\documentclass[lettersize,journal]{IEEEtran}
\usepackage{amsmath,amsfonts}
\usepackage{algorithmic}
\usepackage{algorithm}
\usepackage{array}
\usepackage[caption=false,font=scriptsize,labelfont=sf,textfont=sf]{subfig}
\usepackage{textcomp}
\usepackage{stfloats}
\usepackage{url}
\usepackage{verbatim}
\usepackage{graphicx}
\usepackage{cite}
\usepackage[table,xcdraw]{xcolor}
\usepackage{cuted} 	\stripsep 0pt
\usepackage{hyperref}
\usepackage{upgreek}
\begin{document}

\title{Automotive Radar Mutual Interference Mitigation Based on Hough Transform in Time-Frequency Domain}

\author{Yanbing Li,~\IEEEmembership{Member,~IEEE,}
		Weichuan Zhang,~\IEEEmembership{Member,~IEEE,}
	and Lianying Ji

\thanks{This work was supported by the Fundamental Research Funds for the Central Universities 2022RC008.}
\thanks{Yanbing Li is with the School of Electronic and Information Engineering, Beijing Jiaotong University, Beijing, 100044, China (e-mail: ybli1@bjtu.edu.cn). }
\thanks{Weichuan Zhang is with the Institute for Integrated and Intelligent Systems, Griffith University, QLD, Australia. (e-mail: weichuan.zhang@griffith.edu.au).}
\thanks{Lianying Ji is with the Beijing Muniu Linghang Technology Company, Beijing, 100192, China (e-mail: lianying@muniu.tech).}
}
\maketitle

\begin{abstract}
With the development of autonomous driving technology, automotive radar has received unprecedented attention
due to its day-and-night and all-weather working capability. It is worthwhile to note that more and more vehicles are equipped with automotive radars, resulting in mutual interference between radars. The interference reduces radar target detection performance, making perception information unreliable. In this paper, a novel interference mitigation method based on power-weighted Hough transform is proposed for solving the radar mutual interference and improving the safety of autonomous driving systems. Firstly, the frequency modulation characteristics of interference signals and target echo signals are analyzed, and differences between the two signals are introduced. Secondly, based on the straight line detection technique, the power of the mutual interference signal in time-frequency domain is accumulated, and the accurate position of the interference is located. Finally, the target echo is recovered by autoregressive model. Compared with existing state-of-the-art methods, the proposed method has the ability to retain more useful signals after the interference mitigation, and achieve better interference detection robustness under low signal-to-noise ratio conditions. Simulation experiments and real scenario experiments verify the effectiveness of the proposed
method and show its superiority.
\end{abstract}


\section{Introduction}
\IEEEPARstart{R}{adar}, as an environmental sensing technology, has been introduced in more and more civil fields such as automotive radar, traffic radar, and security radar. On one hand, this is due to the development of chip technology, especially millimeter-wave chip technology. These advances have made it possible to reduce radar design cost and difficulty, which allows radar manufacturers to iterate their products rapidly \cite{27,hasch2012millimeter,saponara2019radar}. On the other hand, the trend of intelligence has led to an unprecedented emphasis on perception technology in many aspects of people’s life, which provides necessary and reliable perception data for the post-processing stage.

One of the most representative civilian applications is automotive radar. From low level assisted driving to high level autonomous driving, radars are included as an important sensor in autonomous driving solutions \cite{1,roos2019radar,bilik2019rise}. It is well known that no single sensor has ability to acquire all desired information well in the real-world scenarios of all conditions. Along this way, multi-sensor fusion techniques are increasingly being used for autonomous driving. As one of the three mainstream sensors, i.e., cameras, radars, and lidars, radars have ability to day-and-night and all-weather work, which is not well demonstrated by the other sensors. Meanwhile, radars have advantage in the radial distance and velocity measurement of targets, which is complementary to the information of the other sensors. A typical autonomous driving solution is equipped with seven millimeter-wave radars in a vehicle, which contains one long-range radar for forward looking, two mid-range radars for both forward and rearward looking, and four short-range radars in the four corners for 360 degree coverage \cite{3,sun2020mimo}. This configuration allows radar sensors on a single vehicle to radiate in all directions on roads. In this case, with the development of autonomous driving, the deployment rate of automotive radars will increase rapidly in the future. As a result, influence among radars become inevitable \cite{12}. Interference among radars may lead to target detection degradation and increase the likelihood of target loss in severe cases, which is unacceptable for traffic safety \cite{4}.

Generally, there are two main categories of radar interference. One category is caused by radar devices interfering with each other, and the other category is spoofing attacks performed by jamming devices. The latter is similar to electronic warfare in military applications and is usually introduced in malicious attacking \cite{6899663}. A research on the suppression of malicious jamming such as digital radio frequency memory (DRFM) jamming is discussed in \cite{8759034}. Compared with spoofing attack jamming, the problem of mutual interference between radars is more common in practical scenarios, especially in high-density traffic flow scenarios.

Many research analyzing mutual interference of automotive radars can be found in \cite{5,6,7,8,kim2018peer}. These sources discuss the occurrence probability of mutual interference between radars, calculate the theoretical value of interference power, and illustrate the interference signal in the time domain, the frequency domain and the time-frequency (TF) domain, respectively. These studies deepen our understanding of the mutual interference for automotive radars and indicate that the mutual interference will worsen signal quality and signal-to-noise ratio (SNR), thereby affecting the target detection ability of radars \cite{ayhan2016impact,9127843,schipper2014discussion}.

Methods used for solving the aforementioned issues can be categorized into two groups according to the degree of dependence on radar system architecture. The first group is coupled with the radar system, and its implementation usually requires specific software and hardware architectures. Approaches based on transmit waveforms such as orthogonal noise and phase-coded are proposed in \cite{9,10}. Another waveform optimization approach is proposed in \cite{9545561}. These methods suppress interference based on the special structure of waveforms. Digital beamforming methods based on radar antenna array structure are discussed in \cite{11,bechter2016digital}, in which interference in specific directions can be suppressed by the directivity of a formed beam. Because interference sources and targets are usually in the same or close direction in a traffic scene, the digital beamforming methods face angle resolution challenges. Inspired by the biological behavior of bats, a heuristic frequency hopping technique is introduced in \cite{bechter2016bats}. When interference occurs, a radar with higher frequency shifts its frequency upwards, while a radar with lower frequency shifts its frequency downwards. This strategy has a higher success rate for interference mitigation than random hopping way. Alternately, radar and communication cooperation is employed for solving mutual interference \cite{8943325}. A distributed network protocol that enables the radar and communication systems to work together is designed, then the avoidance of mutual interference among radars can be achieved due to information sharing. The above-mentioned methods can realize interference mitigation by designing specific system functions, which achieves good effect in designated situations. However, these methods require constraints on radar system design, thereby increasing the development cost and difficulty of radar products.

Another group of methods does not customize the radar software and hardware, but uses signal processing techniques, i.e., signal detection and reconstruction, for suppressing interference on the existing radar system architecture, which has good versatility in practice. In terms of the acquisition domain of interference information, these methods can generally be divided into time domain, frequency domain, and TF domain methods. An adaptive noise canceller that uses interference information in negative frequencies to cancel the interference in positive frequencies is proposed in \cite{jin2019automotive}. This is a typical implementation of interference mitigation in the frequency domain. Besides frequency domain methods, most of current interference mitigation methods are implemented in the time domain and the TF domain. Zeroing or adding a raised cosine window for the disturbed part of a received signal is adopted in \cite{nozawa2017anti}. These two ways achieve the attenuation of interference power, yet lose useful signals in the overlapped part with the interference. Wavelet decomposition and denoising is used in \cite{13} for removing the interference. Due to the decomposition characteristics of the wavelet transform, useful signals in the undisturbed components can be well retained. Signal reconstruction by autoregressive (AR) model is proposed in \cite{15}, which has ability to extrapolate useful signals in the interfered part and retrieve more target information than the zeroing and the windowed methods, however, reconstruction quality will be degraded when a interfered segment is wide. Another signal reconstruction method named iterative method with adaptive thresholding (IMAT) is proposed in \cite{bechter2017automotive} for overcoming the signal gap introduce by zeroing. The IMAT method is a sparse reconstruction technique from main frequency components essentially. All the methods mentioned above obtain interference information from the time domain, and suppress the interference accordingly.

More recently, a research in \cite{8} shows that more structural information of interference can be observed in the TF domain. In this case, more differences between the target echo and the interference can be extracted in the TF domain than in the time domain. TF analysis of a received signal in a interference scenario is performed for locating interference time span region in \cite{16}, followed by beat frequencies interpolation for recovering the target echo. Another TF analysis based method is introduced in \cite{14}. Here the interference is located by a constant false alarm rate (CFAR) detector, followed by a reconstruction process by zeroing, amplitude correction, and Burg-based signal extrapolation, respectively. Experimental results demonstrate that the methods based on TF analysis are superior in interference mitigation performance to time domain methods.

Although the existing TF domain methods \cite{16,14} have shown superiority to the time domain methods \cite{nozawa2017anti,13,15,bechter2017automotive} in performance, we still have to resolve whether the characteristic information of the interference in the TF domain is fully exploited. For instance, the CFAR based method \cite{14} detects and suppresses interference in frequency slices along the TF spectrogram, without considering the time-frequency variation characteristics of the interference. In this case, interference detection is based on the ratio of the interference power at a certain point to the noise level. A good interference detection performance can be obtained under high interference-to-noise ratio (INR) conditions. However, when the interference power is weak, e.g., the interferer radar is far from the victim radar, the projection of the interference power onto each frequency slice may not be enough for supporting accurate interference detection in the TF domain. In this way, degraded interference mitigation performance may be occurred in low INR conditions for the CFAR based method. Based on aforementioned facts, our main question are: (1) Is there a joint time and frequency characteristic of the interference in the TF domain? And whether this time-frequency characteristic can be effectively extracted for detecting and mitigating the interference? (2) Can the INR be improved for enhancing the interference detection performance? Focusing on these two questions, our research demonstrates that the interference has obvious joint time and frequency structural characteristics on the TF plane, that is, it appears as a straight line with a large slope. In addition, inspired by the incoherent integration method in radar target detection \cite{9779438,7455266}, the line structure characteristics of the interference can be used to accumulated the interference power in the TF domain, thus achieving good interference detection performance.

In this paper, the mutual interference of frequency modulated continuous wave (FMCW) automotive radars based on the TF domain is discussed, and a robust interference detection and mitigation approach by power-weighted Hough transform is proposed. To the best of our knowledge, so far there is no research that considers the structure information in the TF domain to robustly detect and locate interference, especially in weak interference and low SNR conditions. Compared with the existing interference mitigation methods based on signal processing technology, the contributions of this paper are as follows:

\begin{enumerate}
	\item{The first mutual interference detection method for automotive radar in terms of structure information in the TF domain is proposed. By analyzing interference signals in a radar receiver, we conclude that the interference in baseband has a linear frequency modulation (LFM) characteristic, i.e., it behaves as a straight line in the TF domain. Based on this structure feature, the Hough transform is used to locate the accurate position of the interference in the TF domain.}
	
	\item{For the first time, the way of power accumulation is introduced into the problem of the interference detection in the TF domain. The classical Hough transform is modified for the TF spectrogram of an FMCW radar signal, namely intensity information is introduced into the Hough transform for the power accumulation. After achieving the interference power accumulation in the Hough parameter space, the INR increases, hence improving the stability of the interference detection}
	
	\item{Compared with the interference mitigation methods based on the time domain, the proposed method has the ability to handle the case of multiple interference. Furthermore, the proposed method is also effective when interference duty cycle is large.}
\end{enumerate}

The rest of the paper is organized as follows. Section~\ref{sec:SigModel} introduces the signal models of the FMCW radar signal and the mutual interference. Then an interference mitigation algorithm based on power-weighted Hough transform is presented in Section~\ref{sec:Method}. Numerical simulations and experimental data based results are shown and discussed in Sections \ref{sec:SimExperi} and \ref{sec:RealExperi}, respectively, to evaluate the interference mitigation performance of the proposed method. Finally, Section~\ref{sec:Conclu} concludes this paper.

\section{Linear FMCW Signal Model in Radar Mutual Interference Cases} \label{sec:SigModel}
\subsection{Linear FMCW Signal Model without Interference} \label{sec:SigModelwoIntf}
A LFM signal, also named a chirp signal, is the most common waveform used in a FMCW radar system in real
applications \cite{3,roos2019radar}. Usually, a set of LFM signal sequences is transmitted from a radar antenna to sense environment. The single transmitted LFM signal is
\begin{equation}
\label{eq1}
\begin{aligned}
	s_{t}(t) &=\sqrt{2 P_{t}} \cos [2 \uppi \upvarphi(t)] \\
	&=\sqrt{2 P_{t}} \cos \left[2 \uppi\left(f_{c}+\frac{1}{2} k t\right) t\right],
\end{aligned}
\end{equation}
where $f_{c}$ is the central carrier frequency, $P_{t}$ is the transmitted power, and $k$ is the chirp rate which equals the ratio of the chirp sweep bandwidth $B$ to the chirp sweep time $T$, i.e., $k=\frac{B}{T}$. The frequency of the transmitted signal is
\begin{equation}
	\label{eq3}
\begin{aligned}
	f_{t}(t) &=\frac{d \upvarphi(t)}{d t} \\
	&=f_{c}+k t .
\end{aligned}
\end{equation}
Thus a frequency modulation direction is defined as up-chirp when $k>0$ and down-chirp when $k<0$, respectively. 

An echo scattered by a target contains added amplitude and Doppler information related to the target’s radar cross section (RCS) and velocity, respectively. For a single-target scenario, the power of the target echo related to free space attenuation is
\begin{equation}
	\label{eq5}
	\begin{aligned}
		P_{e}=\frac{P_{t} G^{2} \uplambda^{2} \upsigma}{(4 \uppi)^{3} R^{4}},
	\end{aligned}
\end{equation}
where $\uplambda$ is the wavelength of the transmitted signal, $G$ is the antenna gain on the line of sight (LOS), $\upsigma$ is the target RCS representing the ability to scatter the power of electromagnetic waves, and $R$ is the distance between the radar and the target on the LOS. The target distance causes a delay between the target echo and the radar reference signal, which is
\begin{equation}
	\label{eq6}
	\begin{aligned}
		\uptau=\frac{2(R+v t)}{c},
	\end{aligned}
\end{equation}\\
where $c$ is the light speed, $v$ is the relative velocity between the target and the radar on the LOS which causes the Doppler frequency shift. From (\ref{eq5}) and (\ref{eq6}), the echo with one target is
\begin{equation}
	\label{eq4}
	\begin{aligned}
s_{e}(t)=\sqrt{2 P_{e}} \cos [2 \uppi \upvarphi(t-\uptau)],
\end{aligned}
\end{equation}
When there exists multiple targets, the echo signal is the superposition of these target’s echoes.

\subsection{Linear FMCW Signal Model with Interference} \label{sec:SigModelwIntf}
When there is interference, the target echo and the interference are superimposed, and then received by a receiver antenna. For a single-interference scenario, without loss of generality, assuming a interferer radar has the same radio frequency (RF) and antenna specifications with a victim radar, i.e., the two radars have the same transmitted power $P_{t}$, wavelength $\uplambda$, and antenna gain $G$, then the interference power in the receiver of the victim radar is
\begin{equation}
	\label{eq8}
	P_{i}=\frac{P_{t} G^{2} \uplambda^{2}}{(4 \uppi)^{2} R_{i}^{2}},
\end{equation}
where $R_{i}$ is the distance between the interferer radar and the victim radar on the LOS. It is worthwhile to note that $R_{i}$ will be equal to $R$ when the interferer radar is installed on the target. Accordingly, the interference signal is
\begin{equation}
	\label{eq7}
\begin{aligned}
	s_{i}(t)=&\sqrt{2 P_{i}} \cos \left[2 \uppi \upvarphi_{i}\left(t-\uptau_{i}\right)\right] \\
	=& \sqrt{2 P_{i}} \cos \left[2 \uppi f_{c i}\left(t-\uptau_{i}\right)+\frac{1}{2} k_{i}\left(t-\uptau_{i}\right)^{2}\right],
\end{aligned}
\end{equation}
where $f_{c i}$ and $k_{i}$ are the central carrier frequency and the chirp rate of the interferer radar respectively. $\uptau_{i}$ is the time delay between the interference and the reference signal. When there exists multiple interference radars, the total interference signal is the superposition of each interference represented in (\ref{eq7}). According to (\ref{eq5}) and (\ref{eq8}), the signal-to-interference ratio (SIR) at the victim radar receiver is
\begin{equation}
	\label{eq9}
	\mathrm{SIR}=\frac{P_{e}}{P_{i}}=\frac{R_{i}^{2} \upsigma}{4 \uppi R^{4}}.
\end{equation}
From (\ref{eq4}) and (\ref{eq7}), the total signal received by the radar receiver is
\begin{equation}
	\label{eq10}
s_{r}(t)=s_{e}(t)+s_{i}(t)+g(t),
\end{equation}
where $g(t)$ is the receiver noise. Dechirp processing of the received signal is achieved by using a low noise amplifier (LNA) and mixing with the reference signal. From (\ref{eq1}), (\ref{eq4}), (\ref{eq7}), and (\ref{eq10}), a beat-frequency signal in baseband can be derived as \eqref{eq11}, where $\upphi_{b}$ and $\upphi_{b i}$ are the constant phase terms. Accordingly, the beat frequency introduced by the target is \cite{10}
\begin{figure*}[!t]
		\hrulefill
	\centering
	\begin{equation}	    
	\label{eq11}
	\begin{aligned}
		s_{b}(t)=& s_{t}(t) \cdot s_{r}(t) \\
		=& s_{t}(t) \cdot\left[s_{e}(t)+s_{i}(t)+g(t)\right] \\
		=& 2 \sqrt{P_{t} P_{e}} \cos \left[2 \uppi \left(k \uptau\right) t+\upphi_{b}\right] +2 \sqrt{P_{t} P_{i}} \cos \left[2 \uppi\left(f_{c}+\frac{1}{2} k t-f_{c i}+k_{i} \uptau_{i}-\frac{1}{2} k_{i} t\right) t+\upphi_{b i}\right] +g(t) ,
	\end{aligned}
	\end{equation}
	\hrulefill
\end{figure*}
\begin{equation}
	\label{eq12}
f_{b}=k \uptau,
\end{equation}
and the beat frequency introduced by the interference is
\begin{equation}
	\label{eq13}
f_{b i}=f_{c}-f_{c i}+k_{i} \uptau_{i}+\frac{1}{2}\left(k-k_{i}\right) t ,
\end{equation}
which is a LFM signal. Substituting (\ref{eq12}) and (\ref{eq13}) into (\ref{eq11}), the beat-frequency signal can be rewritten as
\begin{equation}
	\label{eqre11}
s_{b}(t)=A_{b} \cos \left(2 \uppi f_{b} t+\upphi_{b}\right)+A_{b i} \cos \left(2 \uppi f_{b i} t+\upphi_{b i}\right)+g(t),
\end{equation}
where $A_{b}=2 \sqrt{P_{t} P_{e}}$ and $A_{b i}=2 \sqrt{P_{t} P_{i}}$ are the power of the beat-frequency signal for the target and the interference respectively. Thus, the total beat-frequency signal consists of
three parts, namely the target, the interference, and the noise.
\begin{figure}[ht]
	\centering
	\includegraphics[width=3.5in]{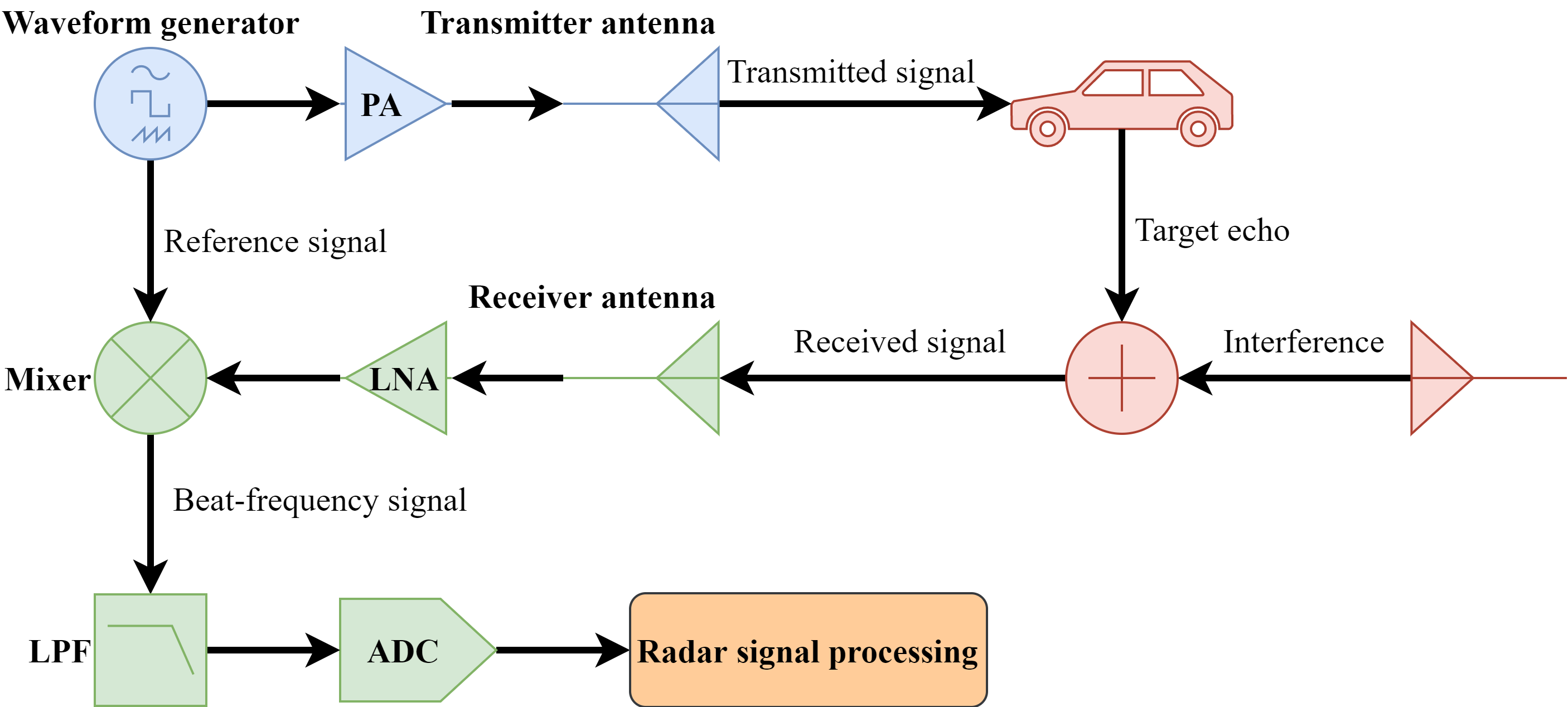}
	\caption{A block diagram of FMCW radar systems.}
	\label{fig_1}
\end{figure}

After the dechirp processing, the beat frequency signal is filtered by a low-pass filter (LPF) whose function is to prevent signal aliasing during subsequent analog-to-digital sampling by an analog-to-digital converter (ADC). Then three fast Fourier transform (FFT) processes, i.e., range FFT, Doppler FFT, and spatial FFT, are applied to the digital signal for estimating the distance, the velocity, and the angle information of the target \cite{3}. A schematic diagram of the FMCW radar system is shown in Fig.~\ref{fig_1}.

\section{Introduction to Interference Mitigation Method} \label{sec:Method}
\subsection{Signal Characteristics and Method Motivation}
Car detection in a typical mutual interference scenario is shown in Table~\ref{tab:tableCarDet}. A car with an interferer radar is present at 100m, and another interferer radar is present at a distance of 2000m. In this case, for a ego radar, SIRs between the car echo and the interference produced by the mounted radar and the distant radar are -41dB and -15dB according to (\ref{eq9}), respectively. These SIR levels indicate the interference power is greater than that of the target echo due to one-way propagation effect shown in (\ref{eq5}) and (\ref{eq8}). As a result, the interferer radar may have an impact on the target detection even if it is far away from the ego radar.
\begin{table}[ht]
	\caption{SIR in a Typical Car Detection Scenario for a Ego Radar in Mutual Interference Cases \label{tab:tableCarDet}}
	\centering
	%	\begin{tabular*}{lc}
		\begin{tabular*}{\hsize}{@{\extracolsep{\fill}}lccc}
			\hline
			Parameters           & Car	&Interferer one	&Interferer two \\ 
			\hline
			Target Distance (m) & 100    & 100	& 2000       \\
			Target RCS at 77GHz ($\mathrm{m}^{2}$) \cite{9266391}      & 10       &-	&-     \\ 
			SIR ($\textbf{dB}$)	&-	&-41	&-15 \\
			\hline
		\end{tabular*}
\end{table}
\begin{figure}[ht]
	\centering
	\includegraphics[width=3.5in]{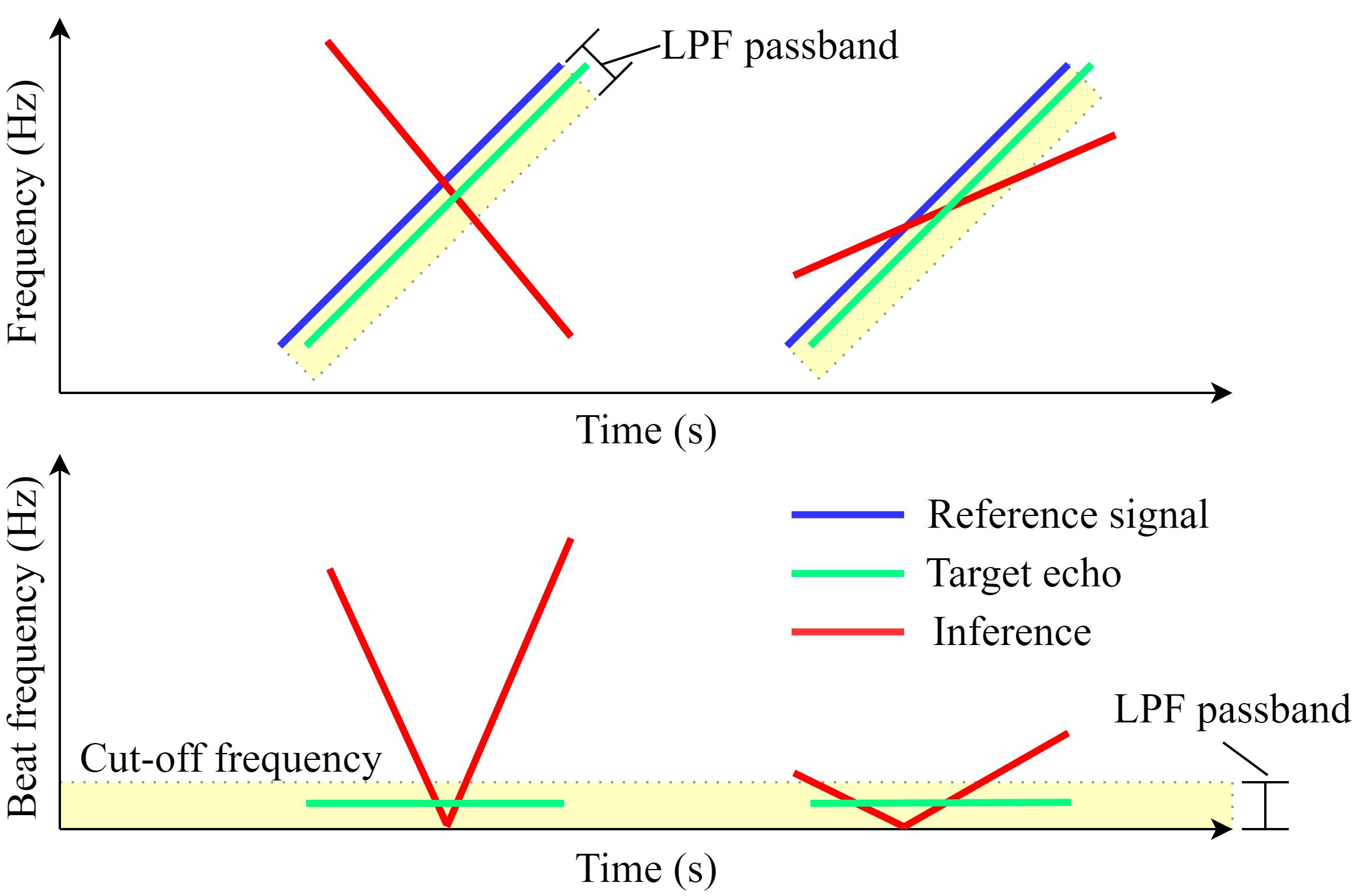}
	\caption{TF characteristics of target echoes and interference before a dechirp process as shown in the first row, and after the dechirp process as shown in the second row. The interference with two frequency modulation directions are given.}
	\label{fig_2}
\end{figure}

In addition, TF features of the target echo and the inference before and after the dechirp processing are shown in Fig.~\ref{fig_2}. As a result of the dechirp processing, it can be seen from (\ref{eq12}), (\ref{eq13}), and (\ref{eqre11}) that the target echo consists of a single-frequency signal, while the interference shows the characteristic of a LFM signal. After low-pass filtering, only signals in passband, which is represented by yellow area in Fig.~\ref{fig_2}, are retained. In this case, the target echo exists in the entire time domain as a single beat frequency signal if its beat frequency is smaller than the cut-off frequency of the LPF. However, since the LFM range of the interference is greater than the LPF passband, the interference will be intercepted by the LPF, which makes the interference exhibits finite range in time, as shown in the second row of Fig.~\ref{fig_2}.

In summary, the target echo and the interference have following characteristics in the automotive radar mutual interference case:
\begin{enumerate}
	\item{The target echo in baseband is a single frequency signal, which demonstrates a straight line parallel to the time axis in the TF domain.}
	\item{The interference in baseband is a LFM and time limited signal, which demonstrates a straight line with a large slope in the TF domain.}
	\item{The interference power is usually greater than that of the target echo due to the difference in signal propagation path. This indicates an automotive radar may be interfered by other radars within a range of kilometers. In this case, the dynamic range of the interference power is large, i.e., both the strong and the weak interferences exist in the received signal of the victim radar.}
\end{enumerate}

Based on the signal characteristic analysis in the automotive radar mutual interference scenario, the dynamic range of the interference power is large in a practical scenario, which brings difficulties to interference detection. The existing interference mitigation methods, such as the wavelet and the CFAR based methods, all take advantage of the larger interference power with respect to that of the target echo for interference
detecting. However, the detection performance of the existing methods decreases with lower interference power. Inspired by the noncoherent integration processing in radar target detection applications \cite{7455266}, the interference detection can be performed by exploiting the line feature of the interference in the TF domain, and accumulating the interference power. Based on this motivation, we propose a Hough transform based interference detection approach in a power accumulation sense. In this way, the interference detection performance can be improved by using the accumulation effect on straight line points in the Hough parameter space.

\subsection{Interference Detection and Localization Based on Power-Weighted Hough Transform}
The characteristics of the interference and the target echo can be obtained by TF analysis technique. STFT is a widely used TF analysis technique due to its good linearity and computational simplicity \cite{17}. In this paper, the TF analysis of the received signal is obtained using STFT and the discrete version implemented in practice is \cite{1163353}
\begin{equation}
	\label{eq14}
S_{r}(\upzeta, m)=\sum_{n=-\infty}^{\infty} s_{r}(n) w(n-\upzeta D) e^{-j 2 \uppi \frac{m}{N} n} ,
\end{equation}
where $w(n)$ is a analysis window (e.g., Hamming window \cite{oppenheim2001discrete}), $N$ is the number of frequency samples, $D$ is the hop size between successive DFTs, and $\upzeta$ denotes the time index. Then, the power spectrogram of the received signal is
\begin{equation}
	\label{eq15}
			\begin{aligned}
P(\upzeta, m)=& \left|S_{r}(\upzeta, m)\right|^{2} \\
=& S_{r}(\upzeta, m) \cdot \text{conj}\left(S_{r}(\upzeta, m)\right),
		\end{aligned}
\end{equation}
where $\text{conj}(\cdot)$ is a conjugate operation. As long as the power spectrogram $P(\upzeta, m)$ is considered as a special image, it can be used as the input of subsequent Hough transform.

As an effective geometric shape detection method \cite{21},\cite{22}, Hough transform has been widely used in many fields such as image processing \cite{23} and lane detection based on radar images \cite{24}. The classical Hough transform detects straight lines in binary images. It projects straight lines into a parameter space to accumulate the scores of points and the straight lines can be obtained by threshold detection in the parameter space. The line function in the Hough transform is defined as
\begin{equation}
	\label{eq16}
\uprho=x \cos (\uptheta)+y \sin (\uptheta) ,
\end{equation}
where $\uprho$ and $\uptheta$ are the distance from the line to the origin and the angle of the line respectively. The coordinate $(x, y)$ is used to describe pixel position for the input image, while each point $(\uprho, \uptheta)$ in the Hough parameter space represents a line in the image. If the line exists in the image, the score of the corresponding point in the parameter space can be measured as
\begin{equation}
		\label{eq17}
		\begin{aligned}
			H(\uprho, \uptheta)=&\sum_{(x,y)} \updelta(x, y) , \\
			\text{with	} 	\updelta(x, y)=&\left\{\begin{array}{l}1, \text { if }(x, y) \text { is on } L \\ 0, \text { otherwise }\end{array}\right. ,
		\end{aligned}
	\end{equation}
where $L$ denotes that the line satisfies (\ref{eq16}). 

Unlike ordinary images, the intensity value of each pixel in the power spectrogram represents the distribution of signal power in the TF domain. From (\ref{eq15}), if a signal has a certain power and chirp characteristics at the same time, it appears as a straight line with the corresponding power in the power
spectrogram $P(\upzeta, m)$. Due to this TF feature, accumulating power information in the Hough parameter space is utilized for improving the performance of the interference detection. The power-weighted score in the Hough parameter space is
\begin{equation}
	\label{eq19}
	H_{P}(\uprho, \uptheta)=\sum_{(\upzeta, m)\in P} P(\upzeta, m) \updelta(\upzeta, m) .
\end{equation}
In addition, considering the slope of the line corresponding to the target echo in the TF spectrogram is close to 0, only lines with large slopes are detected to ensure they correspond to the interference. When the Hough parameter matrix is obtained, the lines can be extracted by threshold detection. With some prior information, the threshold can be determined in a feasible way as follows. In real scenarios, we set a maximum RCS value related to a interested target as $\upsigma_{\max}$ and calculate the theoretical value of the target echo power according to (\ref{eq5}), then the detection threshold is determined as
\begin{equation}
	\label{eq20}
		\begin{aligned}
	\text{Thd}=\frac{\upalpha P_{t} G^{2} \uplambda^{2} \upsigma_{\max }}{(4 \uppi)^{3} R^{4}},
		\end{aligned}
\end{equation}
where $\upalpha$ is the threshold factor which can be determined in a radar test stage. After obtaining the detection threshold, the lines corresponding to the interference can be extracted if $H_{p}(\uprho, \uptheta)>T$, and the interference locations in the spectrogram are found according to (\ref{eq16}) is
\begin{equation}
	\label{eq21}
\left\{\begin{array}{llrl}
	\upzeta & =\uprho, & & \text { if } \sin (\uptheta)=0 \\
	m & =-\cot (\uptheta) \upzeta+\frac{\uprho}{\sin (\uptheta)}, & & \text { otherwise }
\end{array}\right. .
\end{equation}

\subsection{Interference Mitigation and Target Echo Recovery}
Based on the detection results of interference lines, values at interference locations are discarded to achieve interference suppression. Meanwhile, a signal recovery process can be realized by interpolating the discarded locations using neighborhood samples.

For each specific frequency bin slice of the spectrogram, an AR model along the time axis of the spectrogram \cite{chatfield2003analysis} is used for realizing a signal interpolation. The AR model is defined as
\begin{equation}
	\label{eq_ar}
	S_{rec}(\upzeta, m)=\sum_{\upzeta_{n}=1}^{q} \uppsi_{\upzeta_{n}} S_{r}(\upzeta-\upzeta_{n}, m)+\upvarepsilon,
\end{equation}
where $q$ is the number of neighboring samples, $\uppsi$ is the prediction coefficient and $\upvarepsilon$ is the residual. AR coefficients can be obtained by least squares. Therefore, the prediction values are obtained in terms of the solved AR model and the gaps at the corresponding interference locations are filled by the predicted signals for achieving the interference mitigation. After traversing all frequency slices by the interference mitigation process, an interference-free spectrogram is obtained.

Finally, a reconstructed target echo without interference is obtained by applying inverse STFT (ISTFT) to the
interference-free spectrogram. The ISTFT is computed by taking the inverse Fourier transform of each time bin slice of the interference-free spectrogram and overlap-adding the inverted signals \cite{1163353} as
\begin{equation}
	\label{eq_istft}
	s_{rec}(n)=\sum_{\upzeta=-\infty}^{\infty} w_{i}(n-\upzeta D) \frac{1}{N} \sum_{m=0}^{N-1} S_{rec}(\upzeta, m) e^{j 2 \uppi \frac{m}{N} n},
\end{equation}
where $w_{i}(n)$ defines a synthesis window which is the inverse of the analysis window $w(n)$.

In summary, the proposed interference mitigation flow is shown in Algorithm \ref{alg1}.
\begin{algorithm}[ht]
	\caption{Interference Mitigation Method.}
	\begin{algorithmic}
		\STATE 
		\STATE {\textbf{Input:}} received signal $s_{r}$ of one sweep period 
				\STATE {\textbf{Output:}} reconstructed signal $\boldsymbol{s}_{\boldsymbol{\text{rec}}}$ after interference mitigation
				\STATE {\textbf{begin}}
		\STATE \hspace{0.5cm}$S_{r}=\operatorname{\text{STFT}}\left(s_{r}\right)$
		\STATE \hspace{0.5cm}$P=S_{r} \cdot \text{conj}(S_{r})$
		\STATE \hspace{0.5cm}$H_{P}=\text{HoughTransform}(P)$
		\STATE \hspace{0.5cm}$LINES= \text{LineDetection}(H_{P})$
		\STATE \hspace{0.5cm}{\textbf{for}}  $line$ \textbf{in} $LINES$, \textbf{do}
		\STATE \hspace{1cm}$[fbin, tbin]= \text{PositionExtract}( line)$
		\STATE \hspace{1cm}$S_{r}( fbin, tbin )= \text{ARrec}\left(S_{r}( fbin,\overline{tb in})\right)$
		\STATE \hspace{0.5cm}{\textbf{end for}}
		\STATE \hspace{0.5cm}$s_{\text{rec}}=\operatorname{\text{ISTFT}}\left(S_{r}\right)$
		\STATE {\textbf{end}}
	\end{algorithmic}
	\label{alg1}
\end{algorithm}

\section{Numerical Simulation Results} \label{sec:SimExperi}
\subsection{Simulation Description and Evaluation Metrics}
Numerical simulation is one of the two approaches used for evaluating the performance of the proposed interference mitigation method. An FMCW radar signal flow simulation based on Fig.~\ref{fig_1} is carried out, which includes waveform generation, amplification, and emission, free space propagation, target back scattering and interference superposition, low-noise amplification, dechirp processing, lowpass filtering and ADC sampling. Two sampling frequencies are utilized for simulating analog and digital signals respectively, i.e., a large analog frequency (AF) such as 2GHz, is applied for analog signal simulation while an intermediate frequency (IF) is used for analog-to-digital sampling. The main radar parameter settings used in the simulation are shown in Table~\ref{tab:table1}. In the simulation scenario, the interferer radar one and the interferer radar two are set at 30m and 150m from the victim radar respectively.

Two types of targets are set for evaluating different methods as follow:
\begin{enumerate}
	\item{A stationary target is presumed to be located at 150m from the ego radar. It is mainly used for evaluating interference mitigation effects in a single chirp signal in Section~\ref{sec:SimExperi_NoisFree}, and the influence of different SNRs on interference mitigation performance in Section~\ref{sec:SimExperi_SNRs}.}
	\item{A moving target with a speed of 11m/s is presumed to	be located at 100m for evaluating velocity measurement	performance in Section~\ref{sec:SimExperi_MvTar}.}
\end{enumerate}

The performance of the proposed method is compared with seven state-of-the-art methods which included five time domain methods and two TF domain methods. Among them, zeroing, raised cosine window (CW) \cite{nozawa2017anti}, time domain AR (T-AR) \cite{15}, wavelet decomposition \cite{13} and IMAT \cite{bechter2017automotive} methods are implemented in the time domain, STFT beat frequencies interpolation by AR model (STFT-AR) and CFAR-Burg methods are implemented in the TF domain \cite{16,14}. In order to ensure the comparability of each method, a CFAR detector is used for detecting interference positions for all the time domain methods.
\begin{table}[ht]
	\caption{Numerical Simulation Parameters\label{tab:table1}}
	\centering
		\begin{tabular}{lccc}
	\hline
	Parameters                       &Victim &Interferer one &Interferer two \\ \hline
	Operating frequency (GHz)       & 77           & 77                 & 77                 \\
	Sweep bandwidth (MHz)            & 300          & 600                & 600                \\
	Sweep time ($\mu$s)                 & 100          & 10                 & 50                 \\
	Sweep direction                 & Up           & Down               & Up                 \\
	Sampling AF (GHz) & 2            & 2                  & 2                  \\
	Sampling IF (MHz)     & 50           & -                  & -                  \\ \hline                 
	\end{tabular}
\end{table}

The interference mitigation performance in both the time and the frequency domains are evaluated using two time domain metrics and two frequency domain metrics. The first metric is cosine similarity (CS) defined as
\begin{equation}
		\label{eq22}
	\text{CS}=\frac{s_{\text{rec}}^{*} s_{e}}{\left\|s_{\text{rec}}\right\|_{2} \times\left\|s_{e}\right\|_{2}},
\end{equation} 
where $s_{e}$ is the target echo, $s_{\text{rec}}$ is the recovered signal of $s_{e}$, and $*$ denotes the conjugate transposition. The CS is a metric of an angle between two signal vectors, which can be used to represent the correlation of the two vectors. The closer the CS is to 1, the more correlated the two vectors are. The second time domain metric is error vector magnitude (EVM) defined as
\begin{equation}
	\label{eq23}
	\text{EVM}=\frac{\left\|s_{\text{rec}}-s_{e}\right\|_{2}}{\left\|s_{e}\right\|_{2}}.
\end{equation}
The EVM is employed to describe the difference between an ideal signal and a recovered signal. A small EVM value means an accurate reconstruction.

The last two metrics, namely peak sidelobe ratio (PSLR) and integrated sidelobe ratio (ISLR) \cite{7073501,9266564} are employed to evaluate interference mitigation performance in the frequency domain via range profiles. The PSLR is defined as
\begin{equation}
	\label{eq24}
\text{PSLR}=10 \times \log 10\left(\frac{\max_{m\notin [a,b]}F^{2}(m)}{\max_{m\in [a,b]} F^{2}(m)}\right),
\end{equation}
where $F$ is the spectrum of $s_{\text{rec}}$ and the interval $[a, b]$ bounds the main lobe of the spectrum. The PSLR is employed to describe a ratio between the power of the max sidelobes point with respect to that of the main lobe. The ISLR is defined as
\begin{equation}
	\label{eq25}
	 \text{ISLR}=10 \times \log10\left(\frac{\sum_{m=1}^{a} F^{2}(m)+\sum_{m=b}^{M} F^{2}(m)}{\sum_{m=a}^{b} F^{2}(m)}\right),
\end{equation}
The ISLR describes a ratio between the energy of all sidelobes with respect to that of the main lobe. For both the PSLR and the ISLR, smaller values indicate lower sidelobe levels, which represents good interference mitigation performance in our applications.
\begin{figure*}[ht]
	\centering   %  居中
	% 插入图片
	% subfig插入子图
	\subfloat[]{\includegraphics[width=1.65in]{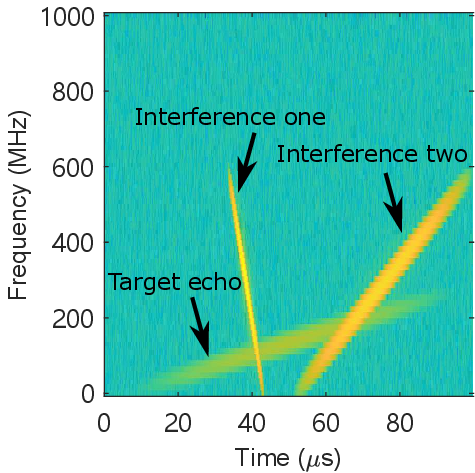}} \hspace{.1in}
	\subfloat[]{\includegraphics[width=1.65in]{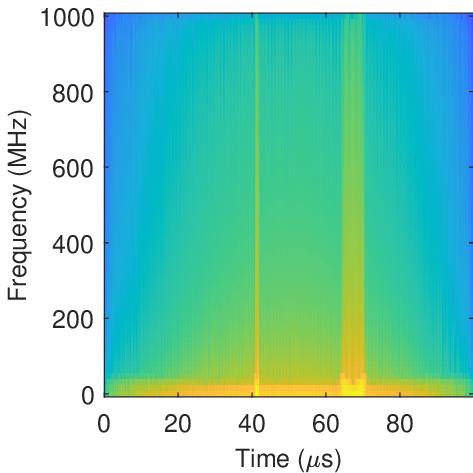}} \hspace{.1in}
	\subfloat[]{\includegraphics[width=1.65in]{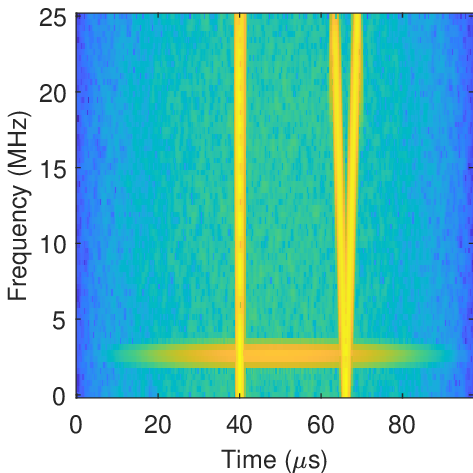}}
	% 图片具体信息
	
	%\includegraphics[width=3.5in]{fig3}
	\caption{Simulated signals. (a) Received signals with target echo and interference in analog domain under the simulation sampling frequency of 2GHz, (b) Signals after LPF in continuous time domain, (c) Signals after ADC under a IF sampling frequency.}
	\label{fig_3}
\end{figure*}
\begin{figure}[ht]
	\centering   %  居中
	% 插入图片
	% subfig插入子图
	\subfloat[]{\includegraphics[width=1.65in]{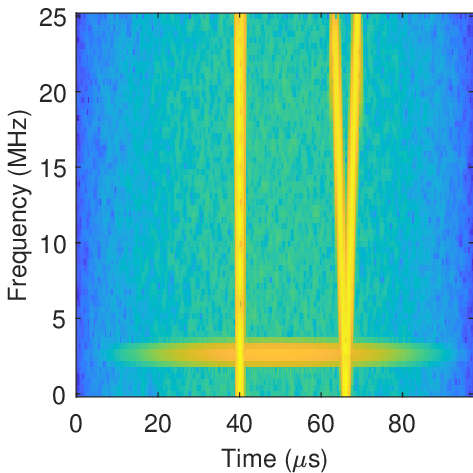}} \hspace{.1in}
	\subfloat[]{\includegraphics[width=1.65in]{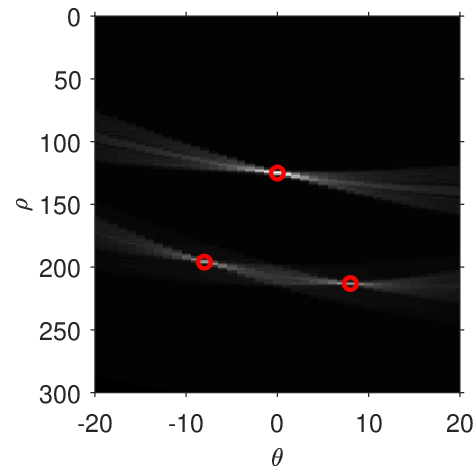}}\\
	\vspace{-0.1in}
	\subfloat[]{\includegraphics[width=1.65in]{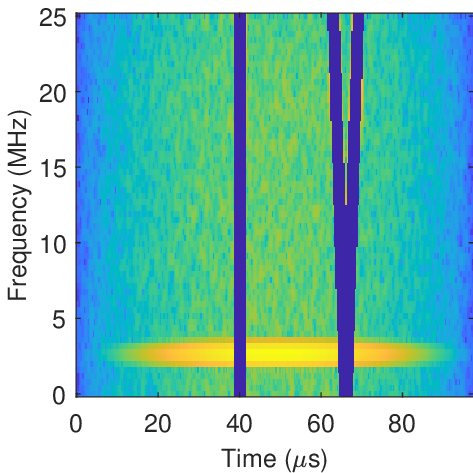}} \hspace{.1in}
	\subfloat[]{\includegraphics[width=1.65in]{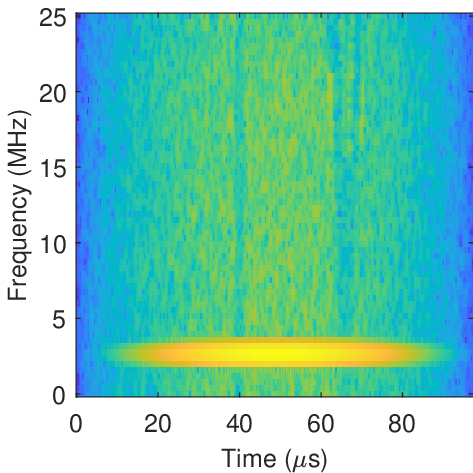}}
	% 图片具体信息
	
	%\includegraphics[width=3.5in]{fig3}
	\caption{Interference mitigation process of the proposed method. (a) The TF power spectrogram of a received signal, (b) Power accumulation results in the Hough parameter space, (c) The Detection of straight lines corresponding to interference in the TF domain, (d) Interference mitigation results in the TF domain.}
	\label{fig_4}
\end{figure}

\subsection{Noise-Free Simulation Results} \label{sec:SimExperi_NoisFree}
Noise-free simulation is used for evaluating the interference mitigation performance firstly. In this case, there are only the target echo and the interference in the received signal, which allows us to quantitatively evaluate the effects of the different interference mitigation methods. The noise-free
signals are shown in Fig.~\ref{fig_3}. The TF distributions of the target echo and two types of the interference in the analog domain are shown in Fig.~\ref{fig_3} (a), the frequencies of the interference and the target echo cross in different time. The interference signals that fell into LPF passband near these intersections are retained, and then received by the radar receiver. The output of LPF and ADC are shown in Fig.~\ref{fig_3} (b) and Fig.~\ref{fig_3}~(c) respectively. 
\begin{table*}[ht]
	\caption{Interference Mitigation Performance in the Noise-free Simulation \label{tab:table2}}
	\centering
	\begin{tabular*}{\hsize}{@{\extracolsep{\fill}}lcrrrr}
		\hline
		Method name                        & Method class & CS                              & EVM                             & PSLR                              & ISLR                              \\ \hline
		Ground truth                           & -  & 1                           & 0                           & -47.151                           & -26.974                             \\
		Zeroing \cite{nozawa2017anti}                           & Time domain  & 0.620                           & 1.191                           & -16.705                           & 1.914                             \\
		CW \cite{nozawa2017anti}                                & Time domain  & 0.630                           & 1.156                           & -16.790                           & 1.659                             \\
		T-AR \cite{15}                               & Time domain  & 0.606                           & 1.210                           & -8.369                            & 2.510                             \\
		IMAT \cite{bechter2017automotive}                               & Time domain  & 0.596                           & 1.348                           & -13.142                           & 2.589                             \\
		Wavelet \cite{13}                            & Time domain  & 0.840                           & 0.597                           & -16.427                           & -4.233                            \\
		STFT-AR \cite{16}                            & TF domain    & 0.998                           & 0.066                           & -37.814                           & -25.805                           \\
		CFAR-Burg \cite{14}                         & TF domain    & 0.998                           & 0.062                           & -39.130                           & -26.006                           \\
		\textbf{Proposed} & \textbf{TF domain}    & \textbf{0.998} & \textbf{0.062} & \textbf{-43.349} & \textbf{-26.008} \\ \hline
	\end{tabular*}
\end{table*}

The interference mitigation effects of the proposed method are shown in Fig.~\ref{fig_4}. The spectrogram of the received signal with interference is shown in Fig.~\ref{fig_4} (a). The power accumulation and the peak detection result in the Hough parameter space is demonstrated in Fig.~\ref{fig_4} (b). Three peaks corresponding to interference lines are detected, and the interference lines in the TF domain are well indicated as shown in Fig.~\ref{fig_4}~(c). Based on these locations, the interference is finely mitigated by the AR model reconstruction process. In this process, the order of the AR model is determined by Akaike information criterion \cite{25,26}. The interference mitigation result in the TF domain is shown in Fig.~\ref{fig_4} (d). Compared with the spectrogram of the received signal with interference as shown in Fig.~\ref{fig_4} (a), the target echo is retained and the interference contaminated areas are reconstructed effectively after the interference mitigation by the proposed method.
\begin{figure*}[ht]
	\centering   %  居中
	% 插入图片
	% subfig插入子图
	\subfloat[]{\includegraphics[width=2.6in]{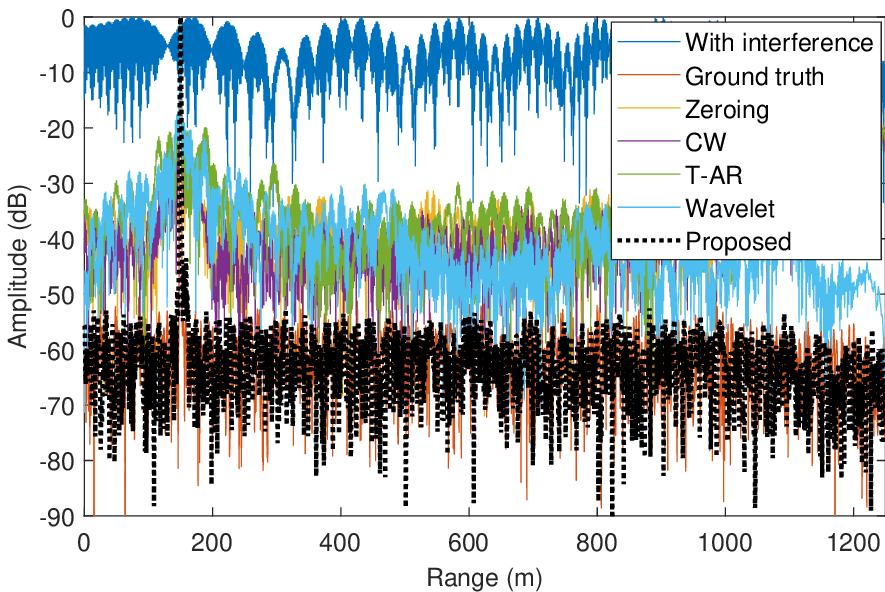}}  \hspace{.1in}
	\subfloat[]{\includegraphics[width=2.6in]{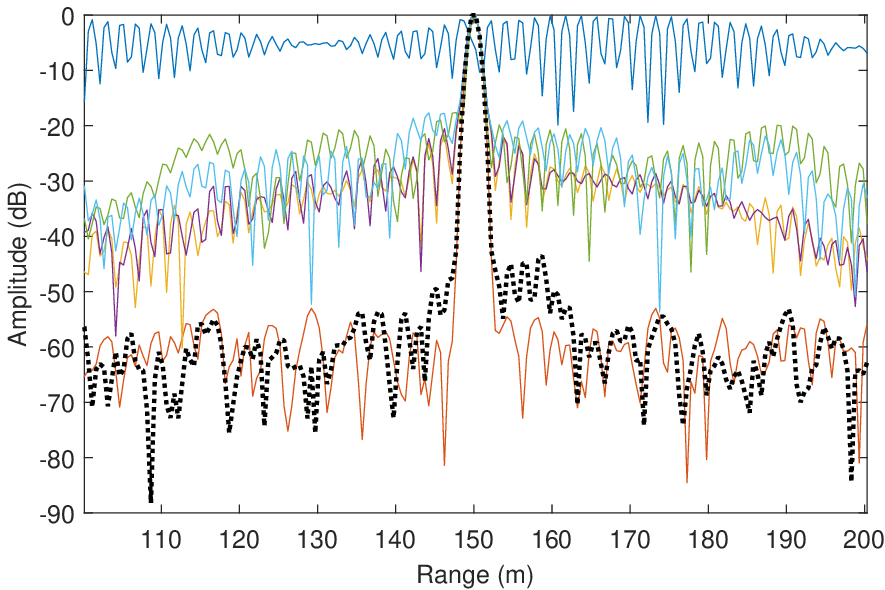}} \\
	\vspace{-0.1in}
	\subfloat[]{\includegraphics[width=2.6in]{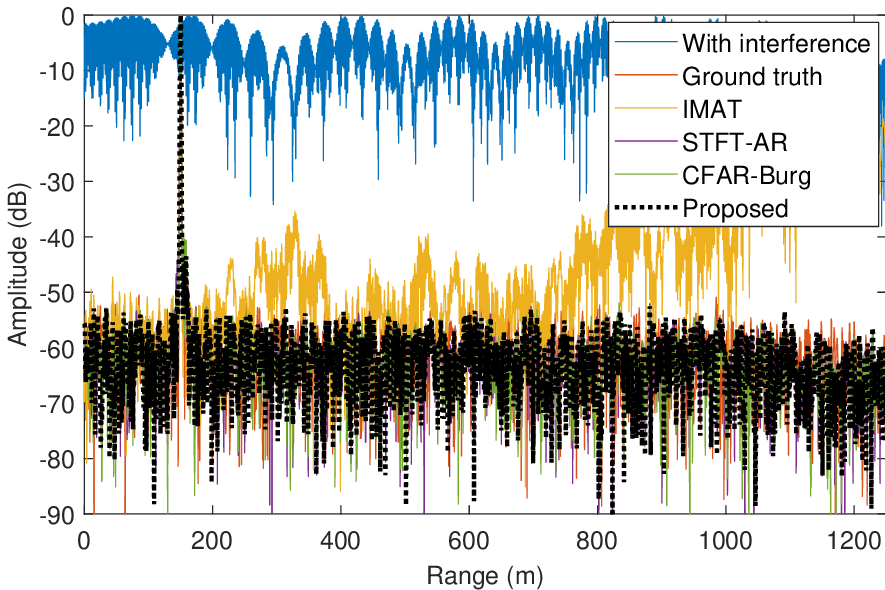}}  \hspace{.1in}
	\subfloat[]{\includegraphics[width=2.6in]{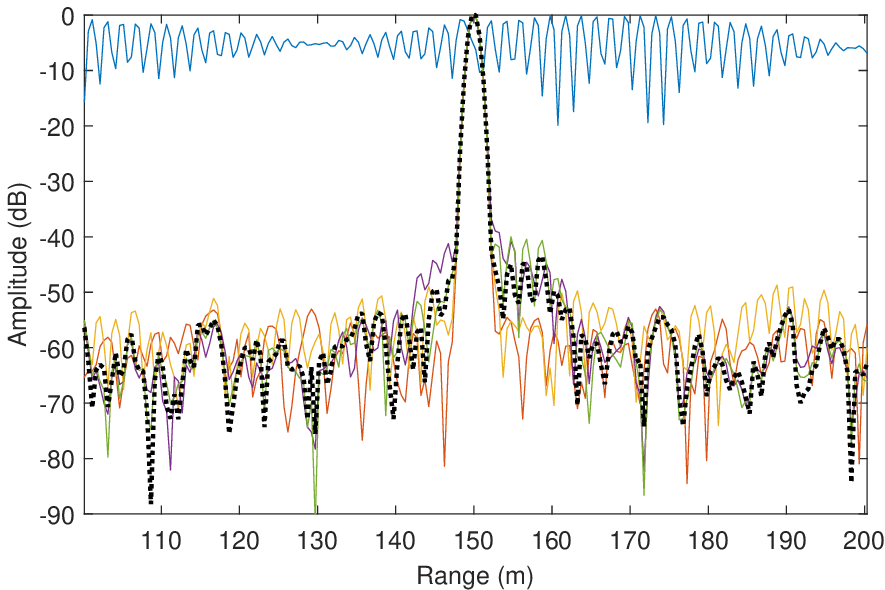}} 
	% 图片具体信息
	
	%\includegraphics[width=3.5in]{fig3}
	\caption{Reconstruction of target echoes in the range profile for simulation data. (a) The first group results, (b) The zoomed-in view of the first group results, (c) The second group results, (d) The zoomed-in view of the second group results.}
	\label{fig_5}
\end{figure*}

The results of the eight methods on the four performance metrics are summarized in Table~\ref{tab:table2}. It can be observed from Table~\ref{tab:table2} that the four time domain methods (i.e., zeroing, CW, T-AR, and IMAT) contain large reconstruction error as shown in the EVM. Furthermore, their correlations with the target echo are poor as shown in the CS. The reason is that the amount of interference information that can be extracted from the time domain is limited. Compared with the four time domain methods, the wavelet method uses the wavelet coefficients of different resolution layers to suppress interference. In this case, the interference power is decomposed into different components, and the useful signals in those components with less interference are preserved. However, the wavelet method still works in the time domain, and does not make full use of the TF characteristics. Unlike the time domain methods, the STFT-AR, the CFAR-Burg, and the proposed methods perform interference mitigation in the TF domain and, therefore, have the ability to exploit more information for accurately locating the interference and retaining more useful signals. Thus the large CS, the small EVM and ISLR are obtained from the three TF domain methods in the noise-free experiment.
\begin{figure*}[ht]
	\centering   %  居中
	\subfloat[]{\includegraphics[width=2.6in]{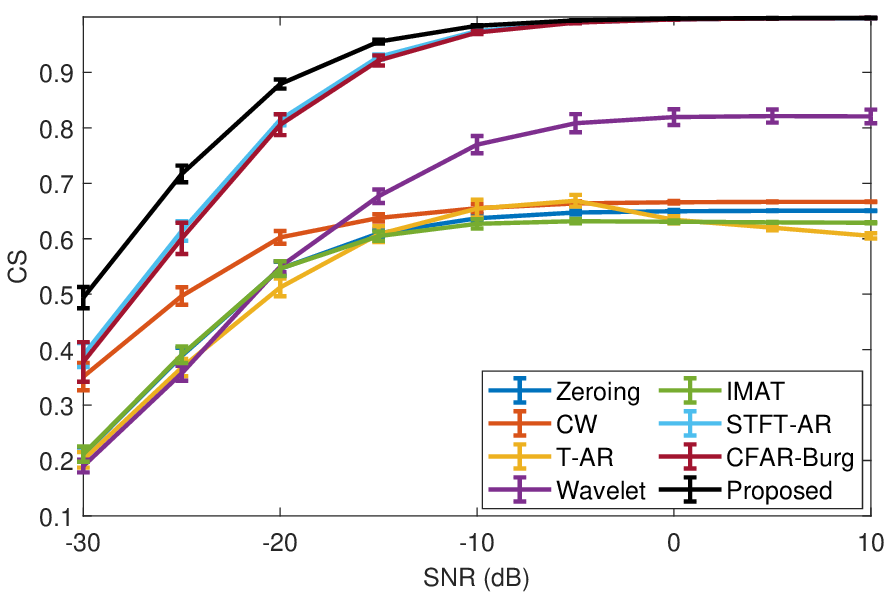}}\hspace{.1in}
	\subfloat[]{\includegraphics[width=2.6in]{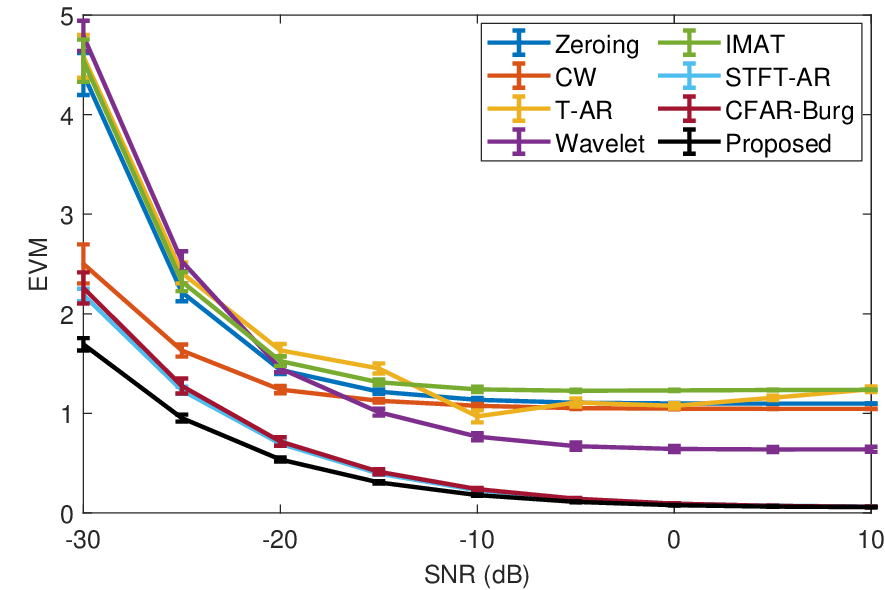}}\\
	\vspace{-0.1in}
	\subfloat[]{\includegraphics[width=2.6in]{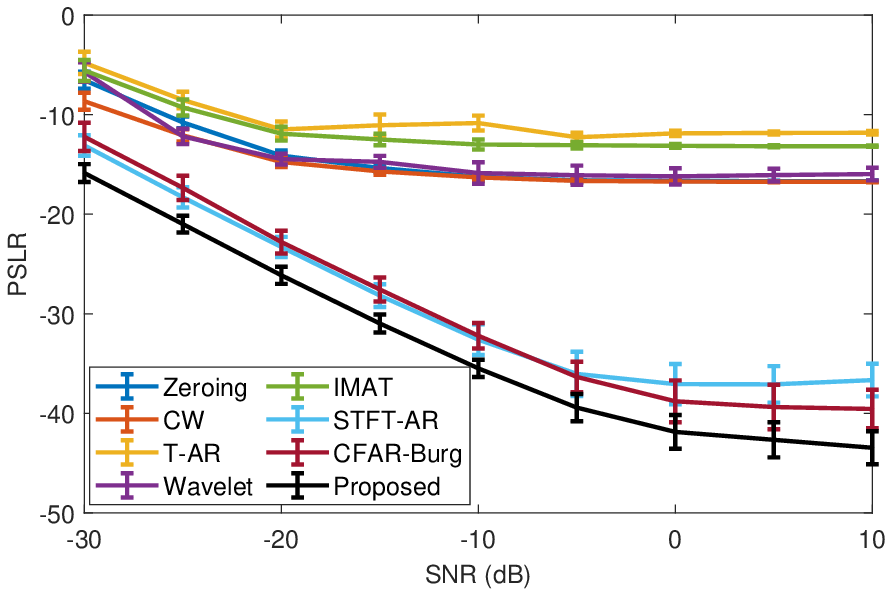}}\hspace{.1in}
	\subfloat[]{\includegraphics[width=2.6in]{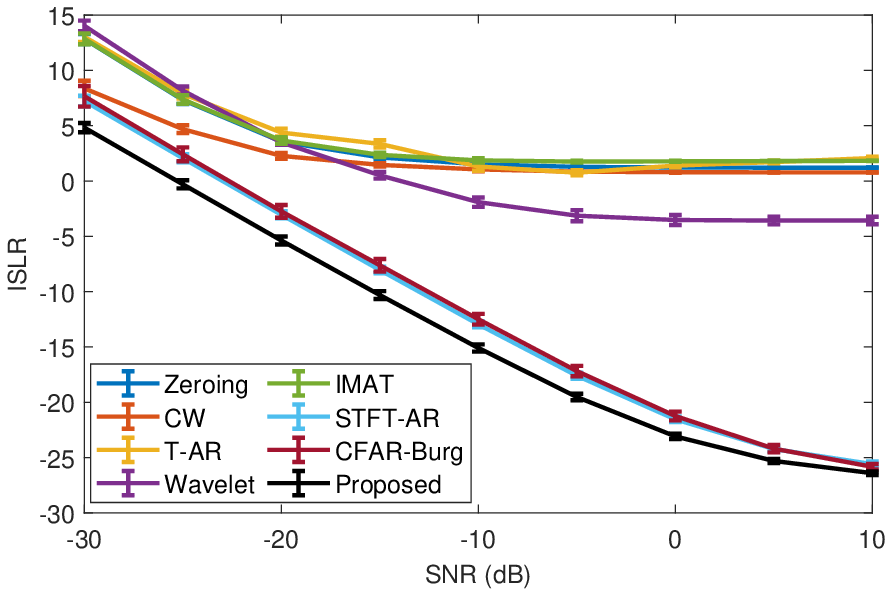}}
	\caption{Interference mitigation performance versus SNR in weak interference simulations. (a) CS, (b) EVM, (c) PSLR, (d) ISLR.}
	\label{fig_6}
\end{figure*}
\begin{figure}[!t]
	\centering   %  居中
	\subfloat[]{\includegraphics[width=1.65in]{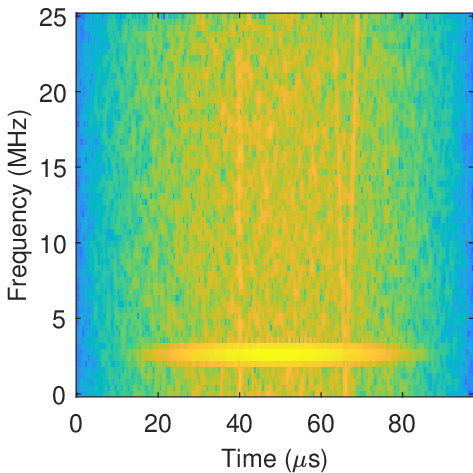}}\hspace{.1in}
	\subfloat[]{\includegraphics[width=1.65in]{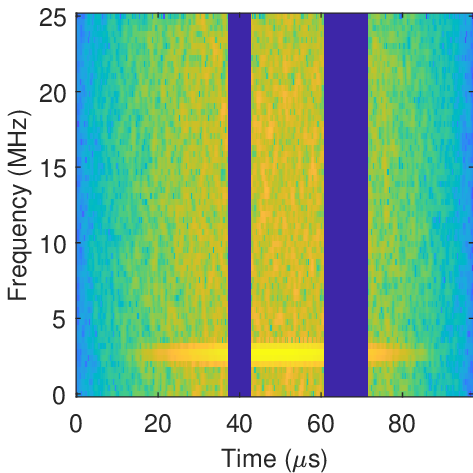}}\\
	\vspace{-0.1in}
	\subfloat[]{\includegraphics[width=1.65in]{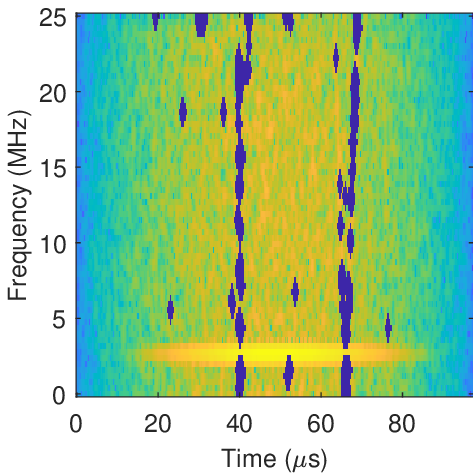}}\hspace{.1in}
	\subfloat[]{\includegraphics[width=1.65in]{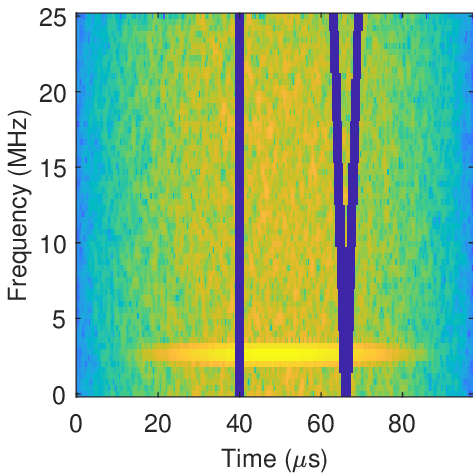}}
	\caption{Interference detection in the TF domain under SNR=-5dB. (a) The TF spectrogram of received signals, (b) Interference removal area for the STFT-AR method, (c) Interference detection by CFAR, (d) Interference detection by the Hough transform.}
	\label{fig_7}
\end{figure}

The frequency spectrum, i.e., the range profile, of the recovered signal are shown in Fig.~\ref{fig_5}. All the eight methods can effectively suppress the interference. Among the five time domain methods, the wavelet method has the best sidelobe levels. Although the IMAT method has the lower sidelobe at the target location of 150m, its sidelobe level deteriorates more rapidly at the distant range. The three TF domain methods have better sidelobe levels than those of the time domain methods and are very close to the signal without interference.
\begin{figure*}[!t]
	\centering   %  居中
	\subfloat[]{\includegraphics[width=2.1in]{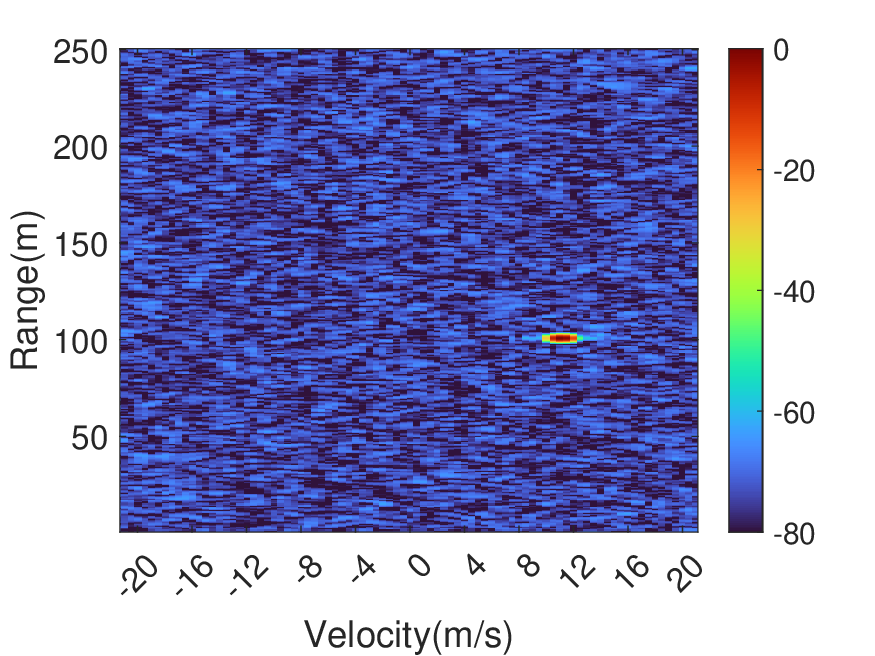}}
	\subfloat[]{\includegraphics[width=2.1in]{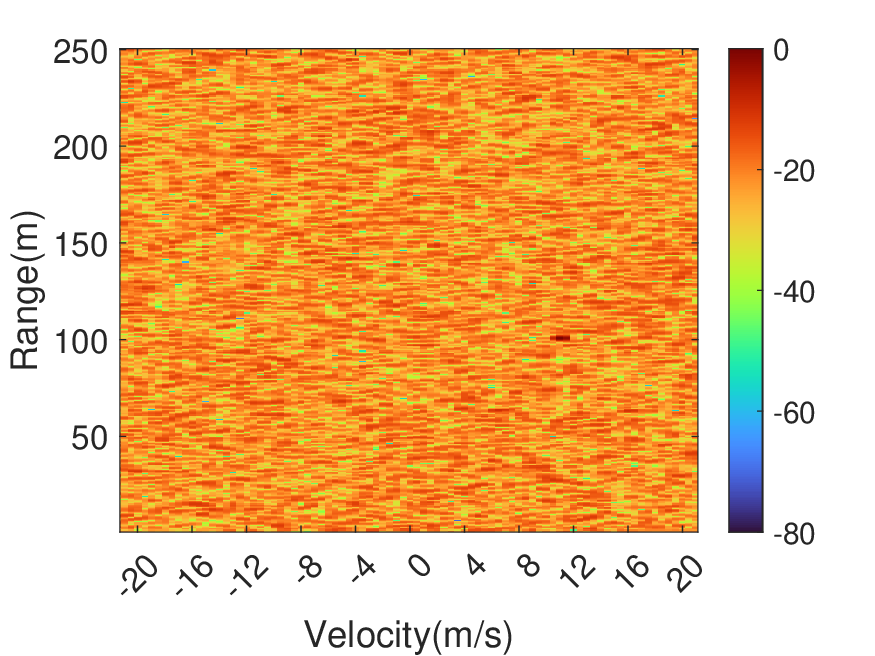}}
	\subfloat[]{\includegraphics[width=2.1in]{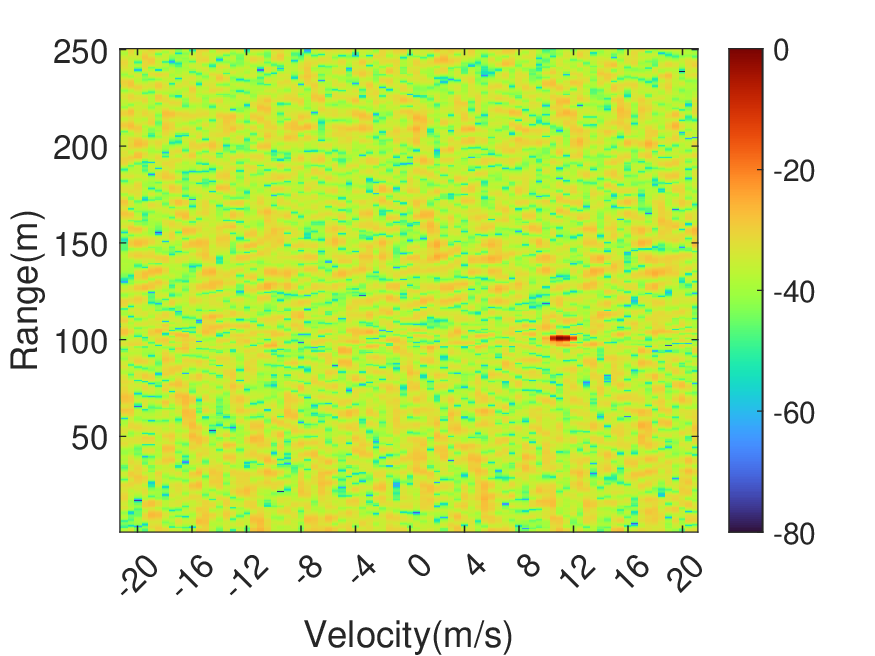}}\\
	\vspace{-0.1in}
	\subfloat[]{\includegraphics[width=2.1in]{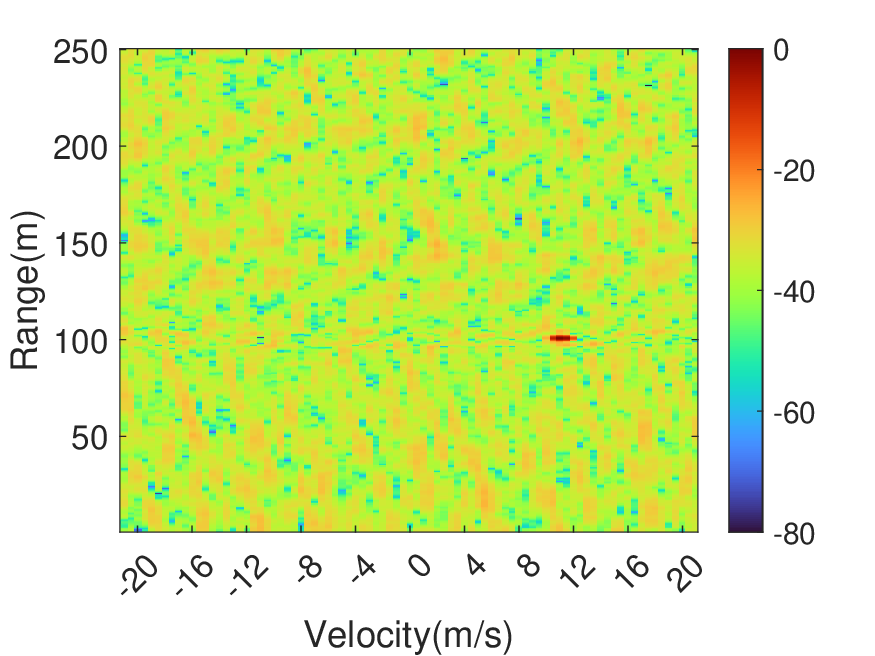}}
	\subfloat[]{\includegraphics[width=2.1in]{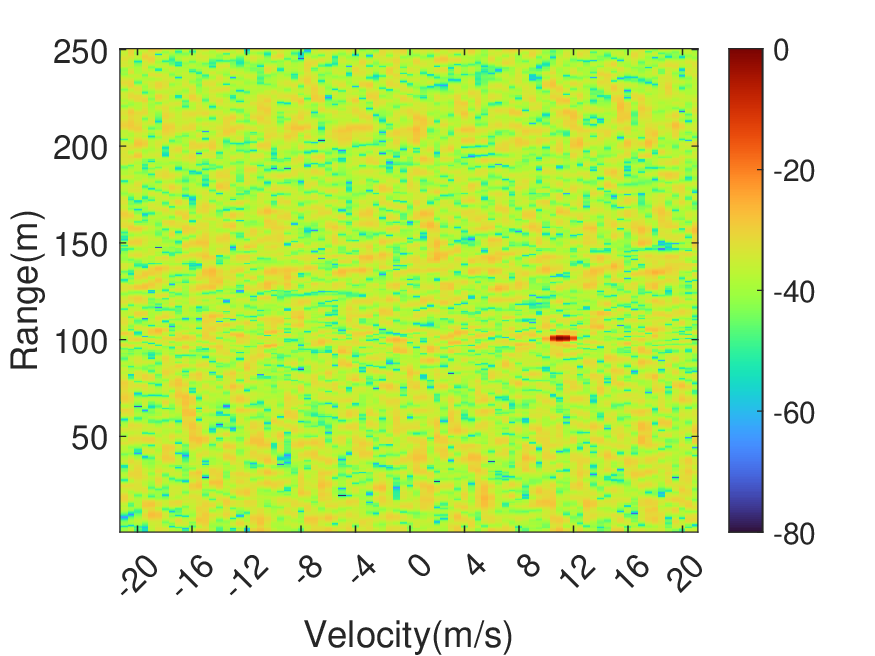}}
	\subfloat[]{\includegraphics[width=2.1in]{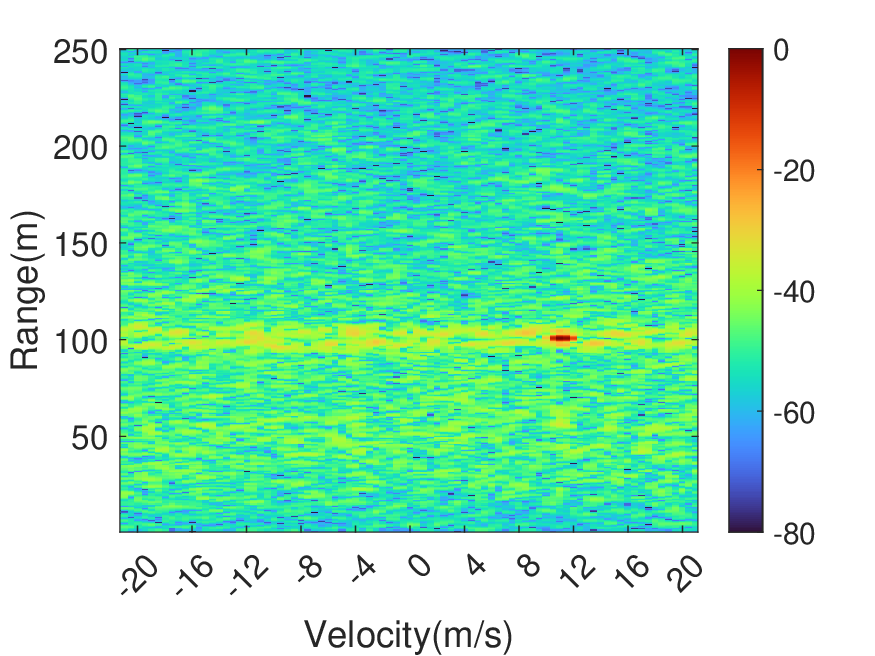}}\\
	\vspace{-0.1in}
	\subfloat[]{\includegraphics[width=2.1in]{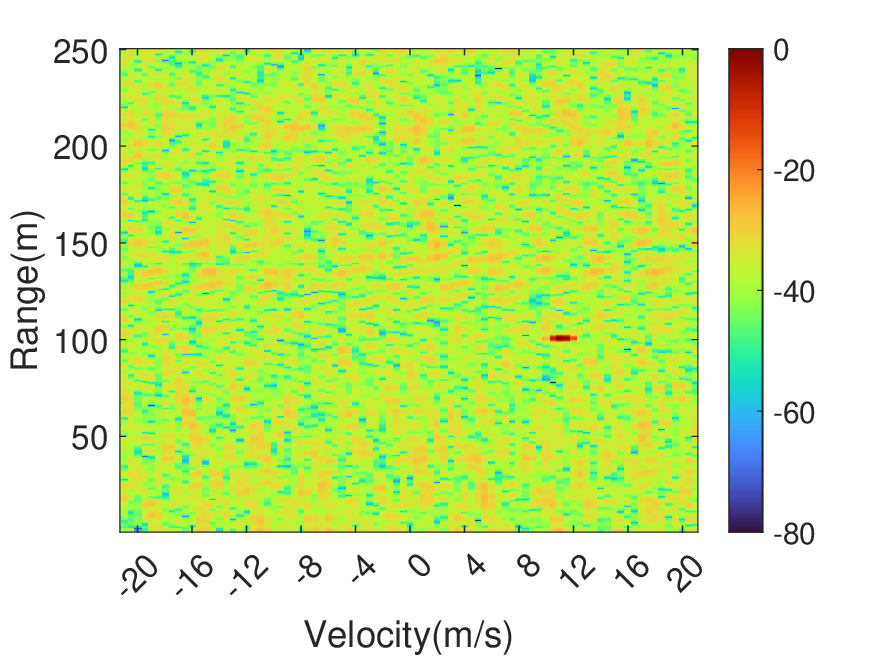}}
	\subfloat[]{\includegraphics[width=2.1in]{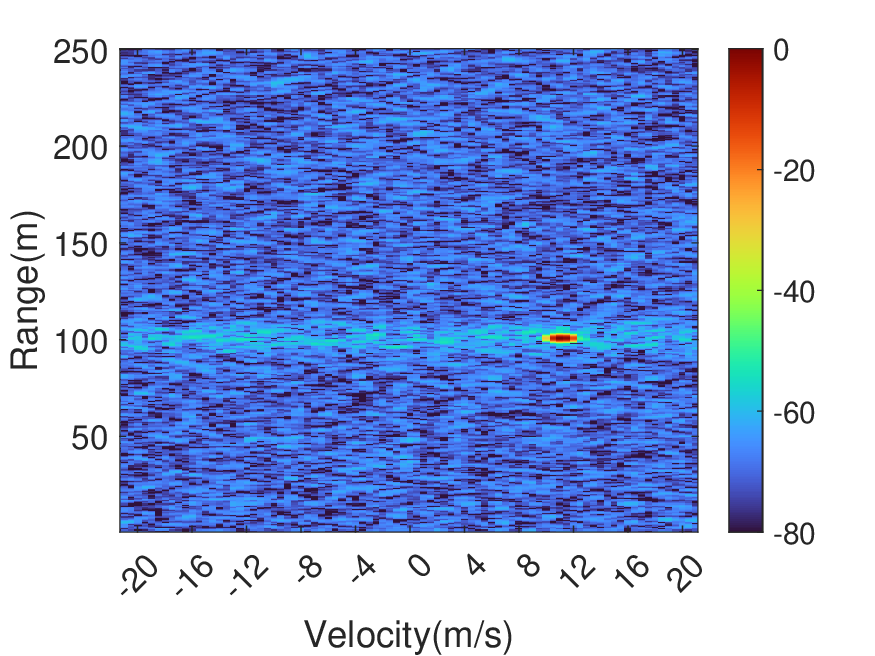}}
	\subfloat[]{\includegraphics[width=2.1in]{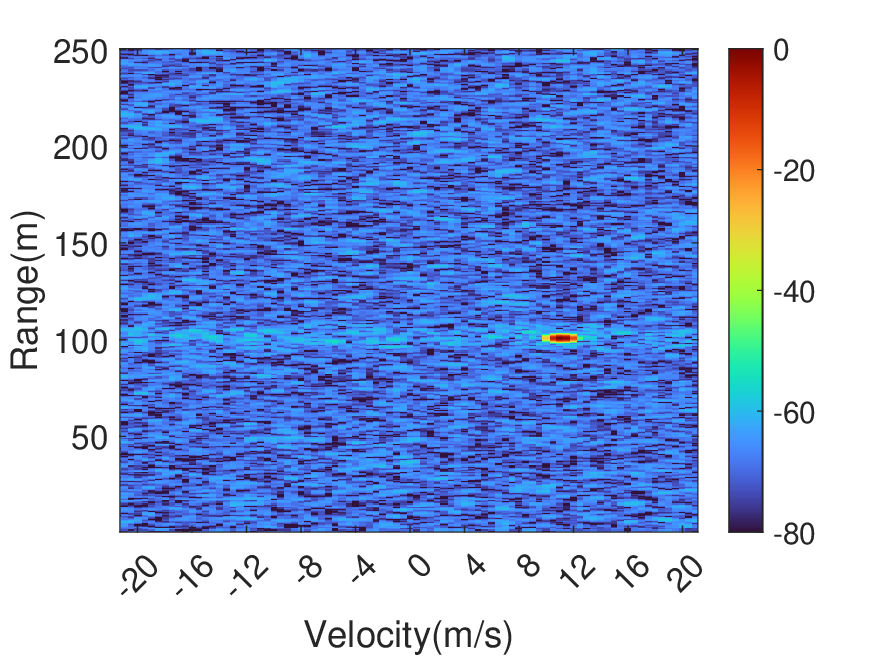}}
	\caption{Zoomed in RD responses before and after interference mitigation in moving target simulations by the tested methods. (a) Ground truth, (b) RD response with interference, (c) Zeroing, (d) CW, (e) T-AR, (f) Wavelet, (g) IMAT, (h) CFAR-Burg, (i) The proposed method.}
	\label{fig_simRd}
\end{figure*}
\begin{figure*}[!t]
	\centering   %  居中
	% 插入图片
	% subfig插入子图
	\subfloat[]{\includegraphics[width=2.6in]{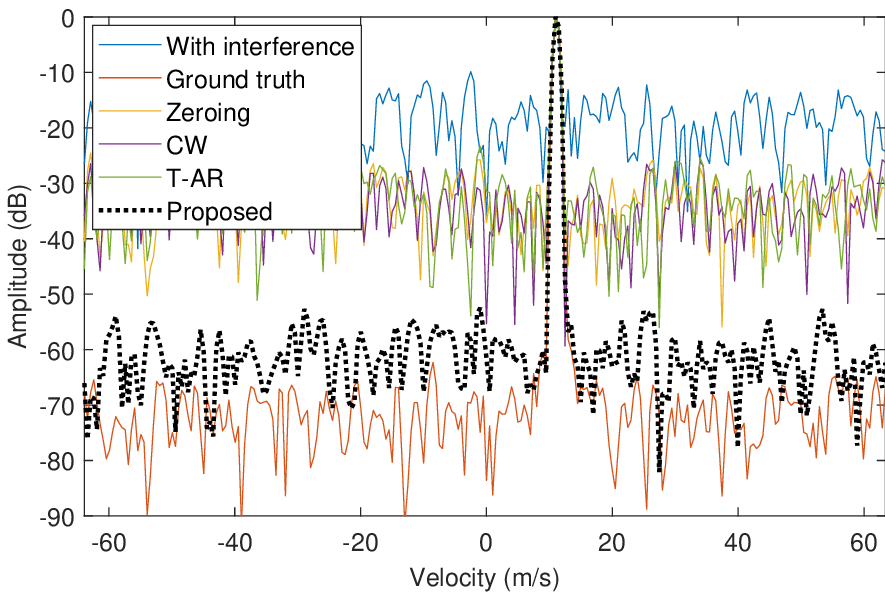}}  \hspace{.1in}
	\subfloat[]{\includegraphics[width=2.6in]{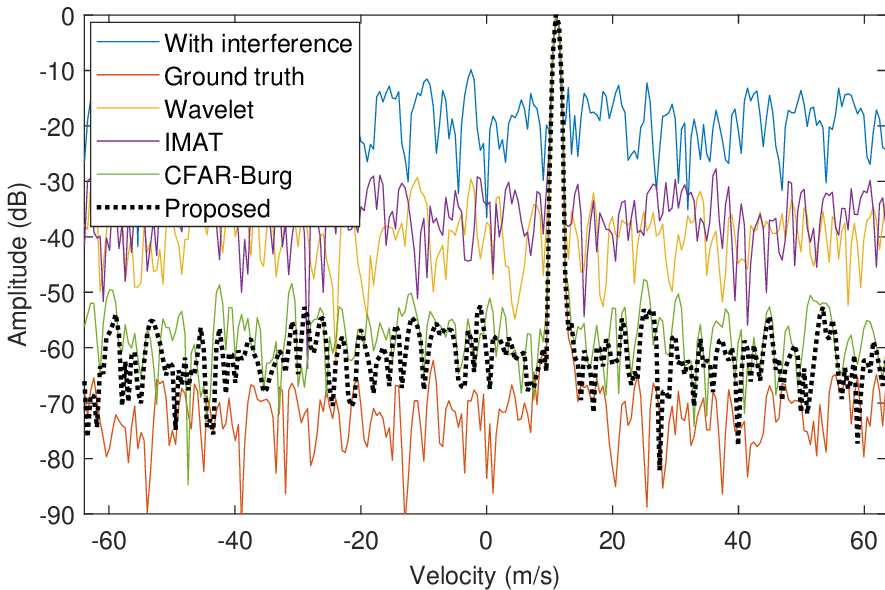}}
	% 图片具体信息
	
	%\includegraphics[width=3.5in]{fig3}
	\caption{The reconstruction of velocity profiles for moving target simulations. (a) The first group results, (b) The second group results.}
	\label{fig_simVel}
\end{figure*}

\subsection{Simulation Results in Different SNR Levels} \label{sec:SimExperi_SNRs}
In this experiment, the interference power is reduced to the same level as the target echo power. This is used for simulating a distant interference scenario and verifying the mitigation performance of the different tested methods for weak interference. Moreover, different SNR level simulations are implemented for evaluating the robust performance of the tested methods. Gaussian white noise with different SNR levels are added into the received signal, and Monte Carlo simulations are repeated for evaluating the statistical performance under specific SNR levels. There are a total of 256 independent noise adding experiments for each SNR level. After all the SNR experiments, the results of the four metrics versus the SNR are shown in Fig.~\ref{fig_6}.

When the SNR is greater than -5dB, for the zeroing, the CW, the T-AR and the IMAT methods, the CS, the EVM, the PSLR and the ISLR of the recovered signal are about 0.8, 0.7, -18, and -3 respectively. It can be found from Fig.~\ref{fig_6} that the wavelet method have achieved better results than the other four time domain methods. Three TF domain methods, namely the STFT-AR, the CFAR-Burg and the proposed methods, achieve the best performance. In this case, the CS is greater than 0.95, the EVM is less than 0.25, the PSLR is less than -32, and the ISLR is less than -15. From these results, the TF domain methods are better than the time domain methods in performance because more information of the interference can be utilized and more accurate interference location can be detected.

When the SNR is low, i.e., smaller than -15dB, the signal recovery performance of all the tested methods is degraded. However, the TF domain methods still maintained advantage over the time domain methods. As shown in Fig.~\ref{fig_6}, in the case of -15dB SNR, the performance of the TF domain methods is still better than that of the time domain methods at high SNR on all the four metrics. In addition, with the decrease of the SNRs, the proposed method has a superiority in the performance and robustness of the interference mitigation to the STFT-AR and the CFAR-Burg methods. For example, when the SNR is -25dB, the CS of the proposed method is about 16\% higher than those of the STFT-AR and the CFAR-Burg methods, and achieves a smaller statistical standard deviation in the Monte Carlo simulations. Similar results are observed on the EVM, the PSLR, and the ISLR as shown in Fig.~\ref{fig_6}.

The interference detection results for the STFT-AR, the CFAR-Burg, and the proposed methods in the TF domain under SNR of -5dB are shown in Fig.~\ref{fig_7}. It can be seen that the proposed method has better interference location detection accuracy than that of the STFT-AR and the CFAR-Burg methods. In low SNR conditions, the power-weighted Hough transform is equivalent to the power accumulation along a straight line in the TF domain. As a result, the INR is improved after Hough transform and the interference is detected robustly.

Unlike the proposed method, there is no accumulation of the interference power to improve the INR in the CFAR-Burg method. Thus it encounters false alarms caused by noise in the low SNR conditions. Moreover, in the frequency slice where the target echo is located, a failure to correctly estimate the noise level lead to a missed detection of the interference for the CFAR-Burg method as shown in Fig.~\ref{fig_7} (c). These corner cases cause the interference mitigation degradation and further affect the signal recovery performance.

As for the STFT-AR method, since there is no explicit interference detection and localization process in \cite{16}, we manually labeled the interference locations for comparison as shown in Fig.~\ref{fig_7} (b). This is an ideal situation. Therefore, the performance in practice will be worse due to detection errors of the interference locations. Compared with the CFAR-Burg and the proposed methods, the STFT-AR method removes all the frequency bins in a certain time range in the TF domain. This operation causes a loss of useful signal information adjacent to the interference locations, and further lead to a decrease in interference mitigation performance.

\subsection{Moving Target Simulation Results} \label{sec:SimExperi_MvTar}
A moving target simulation is used for evaluating the performance of target velocity measurement by chirp sequences before and after interference mitigation. In the simulation, the moving target locates at 100m with a velocity of 11m/s. A total of 256 chirps are set up as a range-Doppler (RD) processing unit. In the evaluation of the interference mitigation performance with multiple chirps, the STFT-AR method is not used because the interference locations vary in each chirp and can not be marked manually.

In the experiment, the interference mitigation process was firstly performed by traversing each chirp to obtain interference-free chirp sequence data, and then the range FFT and the Doppler FFT processing mentioned in Section~\ref{sec:SigModelwIntf} was realized on the interference-free chirp sequence data for obtaining RD responses. The RD responses corresponding to the tested methods are shown in Fig.~\ref{fig_simRd} (c) to Fig.~\ref{fig_simRd} (i). As a reference, the RD responses under the interference-free and the interference conditions are given in Fig.~\ref{fig_simRd} (a) and Fig.~\ref{fig_simRd}~(b), respectively. The CFAR-Burg and the proposed method have better interference mitigation effects in the RD responses than those of the time domain methods, i.e. the zeroing, the CW, the T-AR, the wavelet, and the IMAT methods. The reason is that, on the one hand, the interference can be more accurately located in the TF domain, and on the other hand, the CFAR-Burg and the proposed methods utilize the uncontaminated signals to interpolate the contaminated gaps. Thus the two methods maintain the phase coherence of the chirp sequence, and obtain the high target SNR in the RD responses.

Moreover, compared with the false alarms of the CFAR-Burg method as shown in Fig.~\ref{fig_7} (c), the proposed method utilizes the linear structure of the interference in the TF domain and accumulates the interference power in the Hough parameter space for robustly interference detecting, which is able to avoid the false alarms. Therefore, the proposed method has better performance in reconstructing the signal and obtains a higher SNR in the RD response. Clearer results are given in Fig.~\ref{fig_simVel}. In the RD responses, the range slice of the moving target is extracted for obtaining velocity profiles, then the velocity profiles of the tested methods are shown in Fig.~\ref{fig_simVel}. Similar to the RD response analysis, the proposed method obtains the best interference mitigation performance in the velocity profiles.
\begin{figure}[ht]
	\centering   %  居中
	% 插入图片
	%	\includegraphics[width=1.75in]{fig8}
	%	\includegraphics[width=1.75in]{fig8b}
	% subfig插入子图
	\subfloat[]{\includegraphics[width=1.65in]{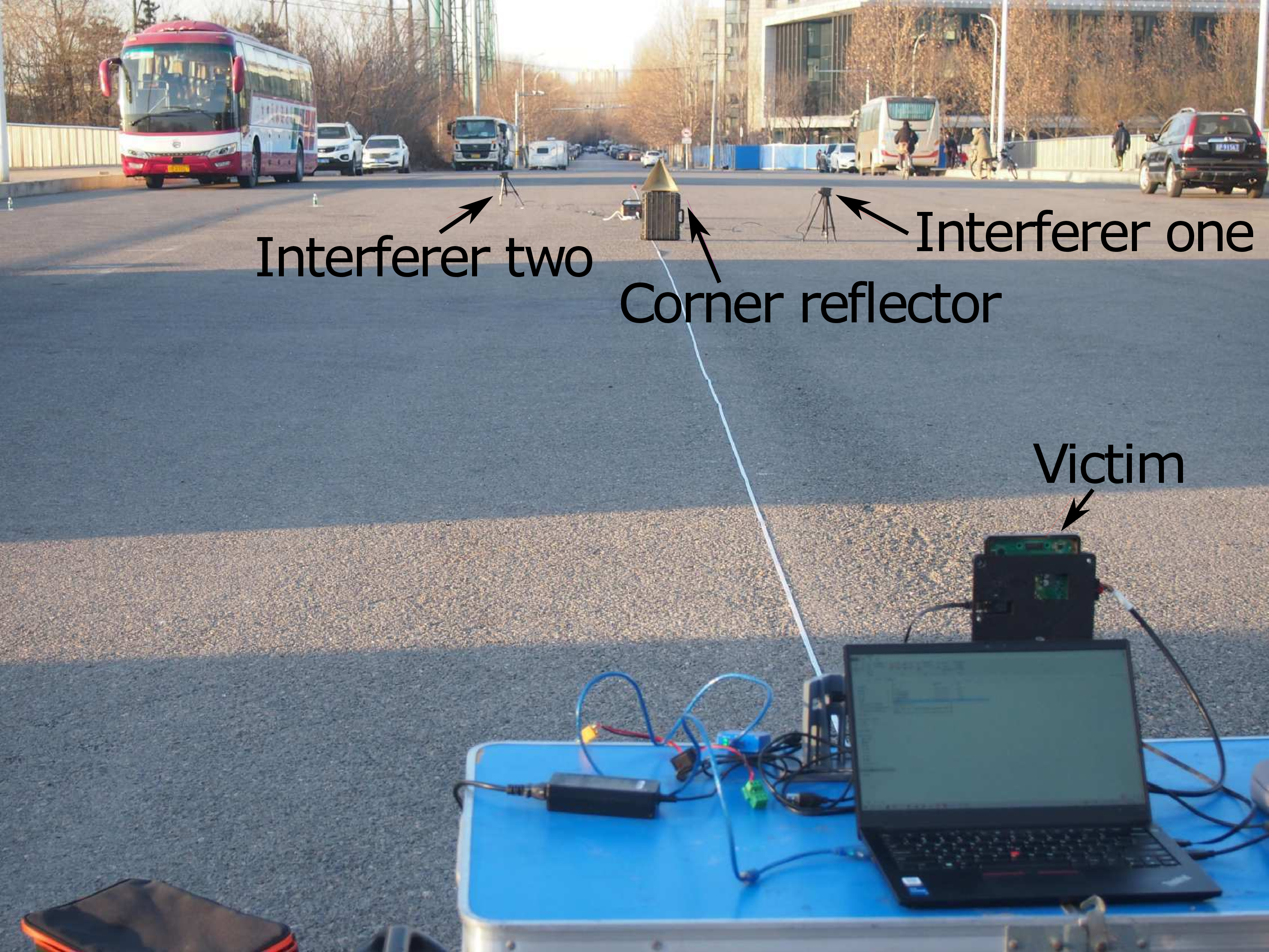}}  \hspace{.1in}
	\subfloat[]{\includegraphics[width=1.65in]{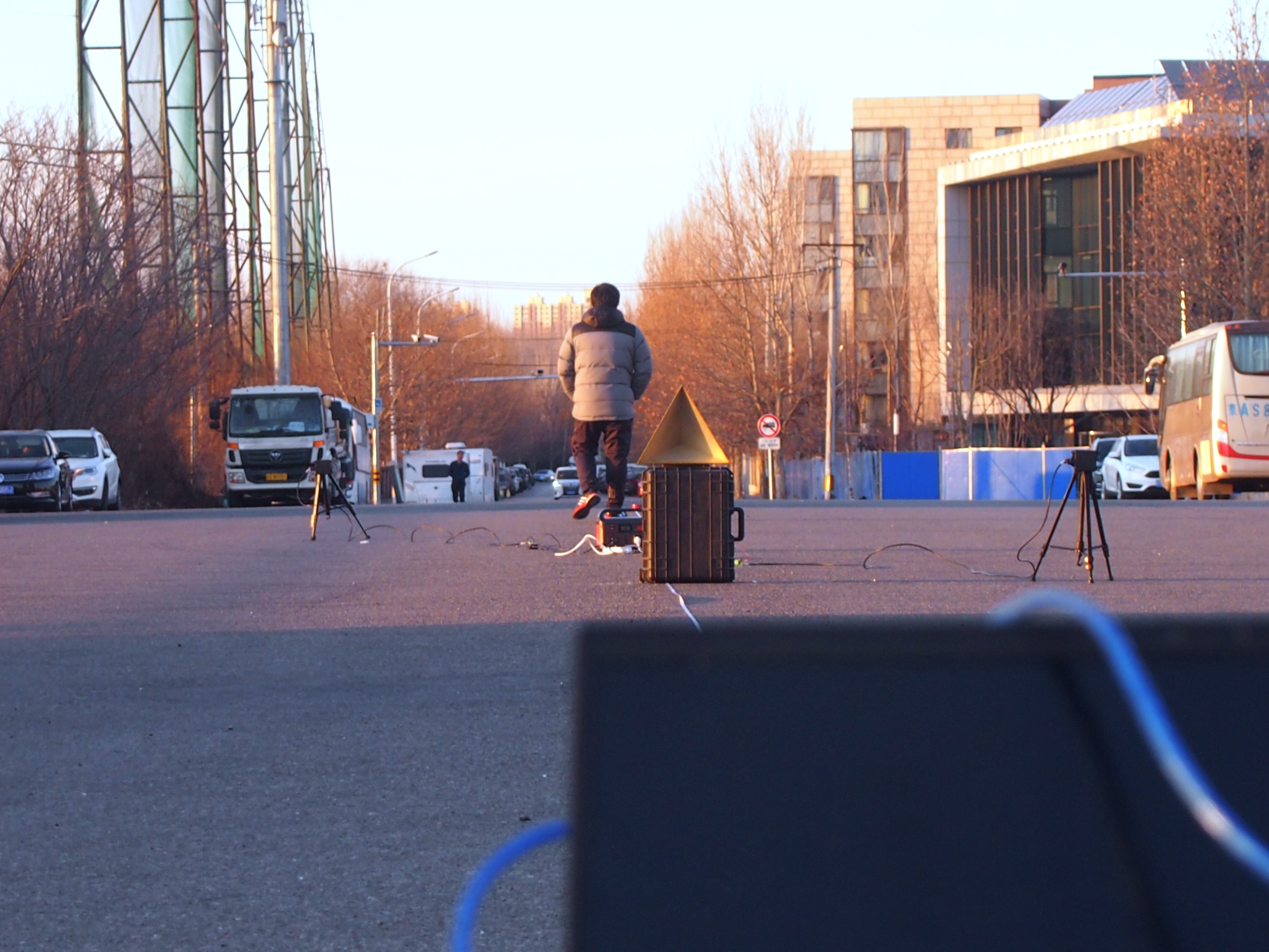}}
	%	\subfloat[]{}
	%	% 图片具体信息
	
	%\includegraphics[width=3.5in]{fig3}
	\caption{Data collection scenario. (a) The schematic of the experimental data acquisition system
		locations, (b) Pedestrian data collection.}
	\label{fig_8}
\end{figure}
\begin{figure}[ht]
	\centering   %  居中
	% 插入图片
	% subfig插入子图
	\subfloat[]{\includegraphics[width=1.65in]{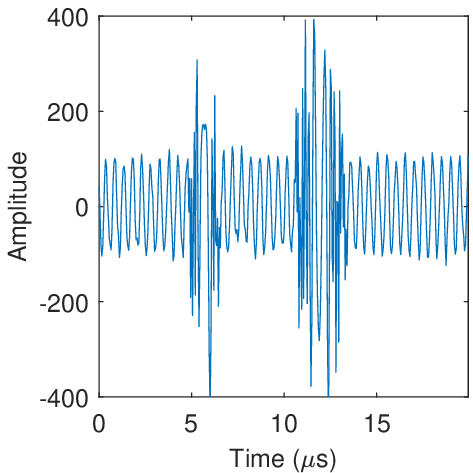}} \hspace{.1in}
	\subfloat[]{\includegraphics[width=1.65in]{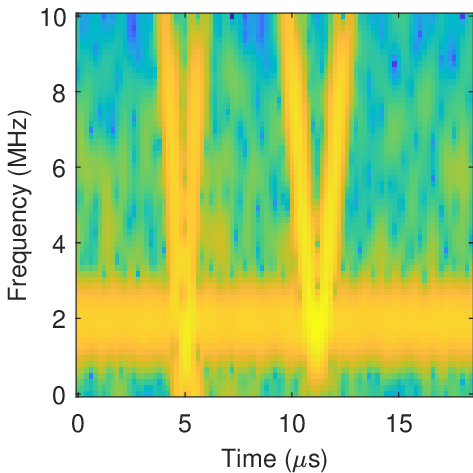}}
	% 图片具体信息
	
	%\includegraphics[width=3.5in]{fig3}
	\caption{A received signal in the real scenario. (a) The time-domain signal, (b) The TF spectrogram.}
	\label{fig_9}
\end{figure}

Quantitative results are given in Table~\ref{tab:MovingTar}. Since the velocity profile is the output of the Doppler FFT processing, the frequency domain metrics, i.e., the PSLR and the ISLR are used to measure the interference mitigation performance for the tested methods. The PSLRs of the time domain methods are distributed between -30dB and -23dB. Among them, the zeroing, the CW and the T-AR methods have the PSLRs around -25dB. These three methods only perform interference detection in the time domain which has the least amount of interference information. The wavelet method decomposes the time domain signal into components and performs interference detection in the components which can retain some useful signal information. As a result, the wavelet method achieves the PSLR level of -29 dB, which is the best performance among the time domain methods. The IMAT method is a sparse reconstruction method for the time domain signal, and this sparse approach makes it to obtain the PSLR level of -27 dB in the velocity profile. Compared with the time domain methods, the TF domain methods significantly improve the PSLR level. The CFAR-Burg and the proposed methods obtain the PSLR levels of -47dB and -52dB, respectively. Since the false alarms occurred in the CFAR-Burg method are avoided in the proposed method, the proposed method obtains the closest PSLR level to the ground truth. Similar results are obtained in the ISLR metric as shown in Table~\ref{tab:MovingTar}.
\begin{figure*}[!t]
	\centering   %  居中
	% 插入图片
	% subfig插入子图
	\subfloat[]{\includegraphics[width=2.6in]{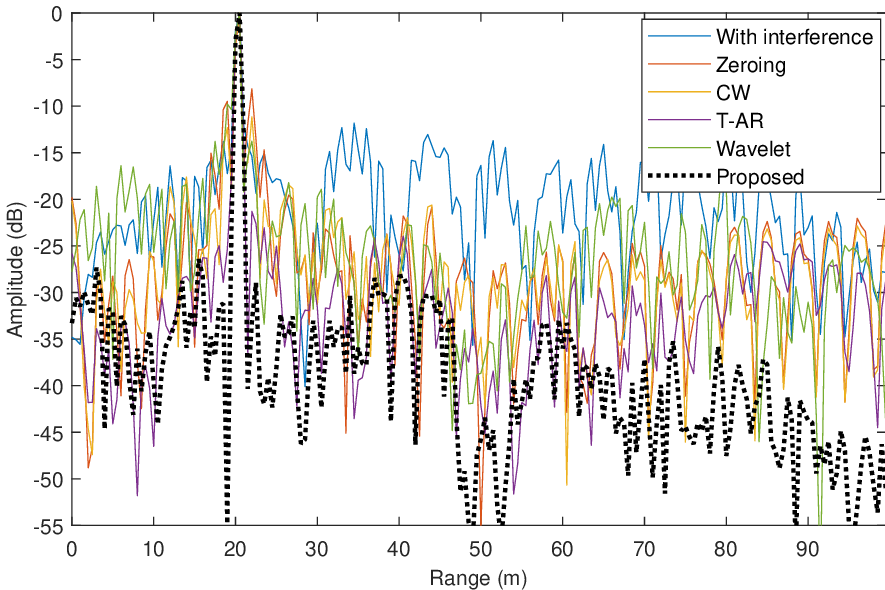}} \hspace{.1in}
	\subfloat[]{\includegraphics[width=2.6in]{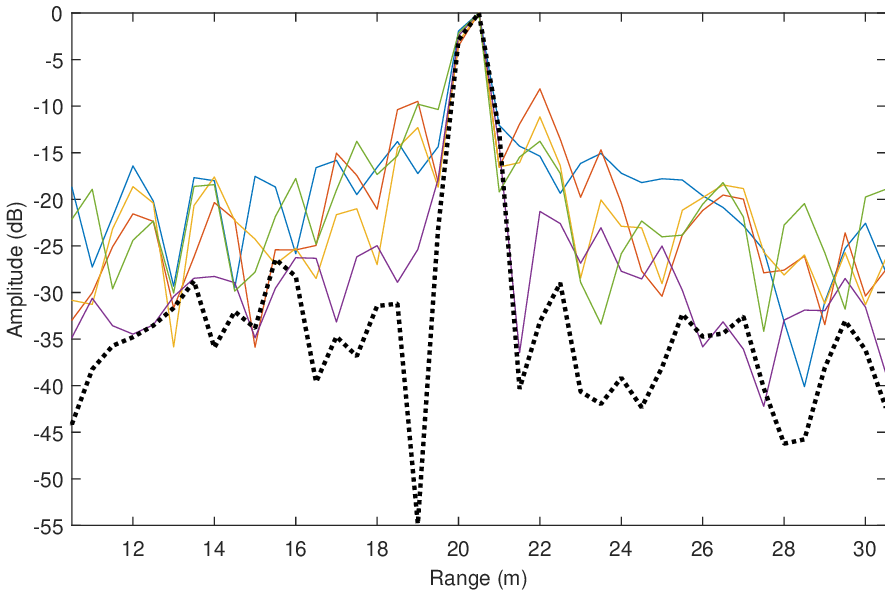}} \\
	\vspace{-0.1in}
	\subfloat[]{\includegraphics[width=2.6in]{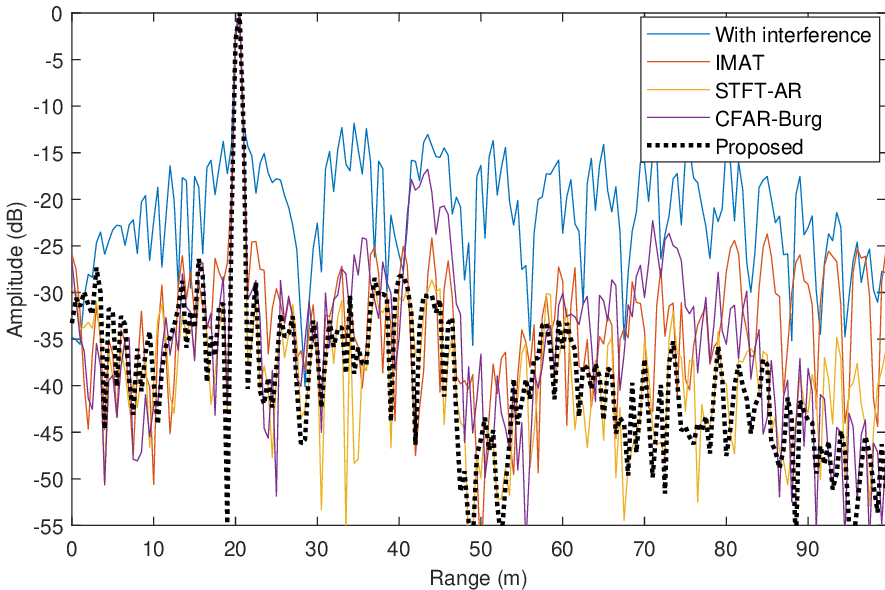}} \hspace{.1in}
	\subfloat[]{\includegraphics[width=2.6in]{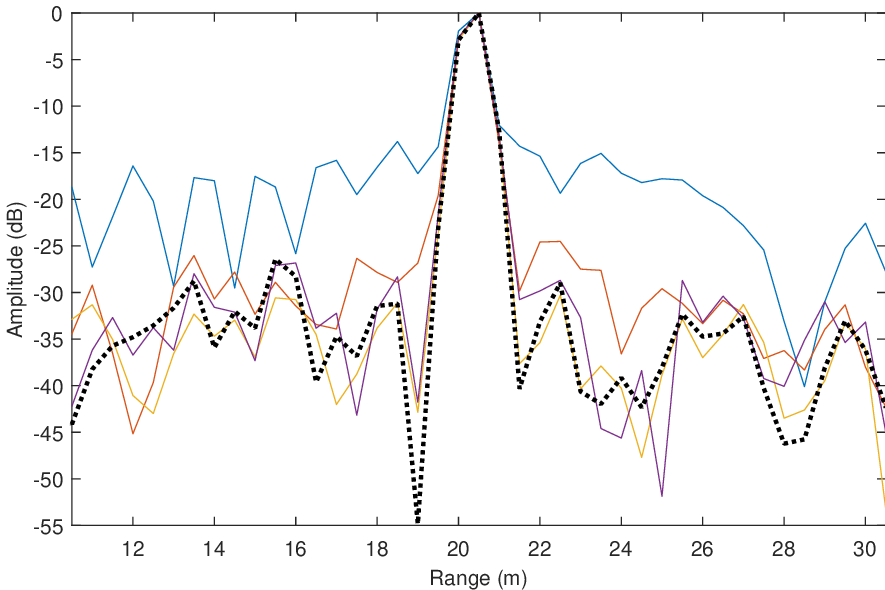}}
	% 图片具体信息
	%\includegraphics[width=3.5in]{fig3}
	\caption{The reconstruction of target echoes in the range profile for experimental data. (a) The first group results, (b) The zoomed-in view of the first group results, (c) The second group results, (d) The zoomed-in view of the second group results.}
	\label{fig_10}
\end{figure*}
\begin{table}[ht]
	\caption{Velocity Profile Reconstruction Performance in Moving Target Simulations \label{tab:MovingTar}}
	\centering
	\begin{tabular*}{\hsize}{@{\extracolsep{\fill}}lccc}
		\hline
		Method name                        & Method class  & PSLR                              & ISLR                              \\ \hline
		Ground truth                           & -   & -60.193                           & -48.986                             \\
		Zeroing \cite{nozawa2017anti}                           & Time domain   & -24.549                           & -10.536                             \\
		CW \cite{nozawa2017anti}                                & Time domain   & -25.760                           & -11.173                            \\
		T-AR \cite{15}                               & Time domain   & -23.318                            & -10.337                             \\
		IMAT \cite{bechter2017automotive}                               & Time domain   & -27.735                           & -13.297                             \\
		Wavelet \cite{13}                            & Time domain   & -29.315                          & -16.081                           \\
		CFAR-Burg \cite{14}                         & TF domain     & -47.715                           & -34.850                           \\
		\textbf{Proposed} & \textbf{TF domain}     & \textbf{-52.459} & \textbf{-38.637} \\ \hline
	\end{tabular*}
\end{table}
\begin{table}[!t]
	\caption{Radar Configurations for Data Collection\label{tab:table3}}
	\centering
	\begin{tabular}{lccc}
		\hline
		Parameters          & Victim & Interferer one & Interferer two \\ \hline
		Operating frequency (GHz) & 77     & 77          & 77          \\
		Sweep bandwidth (MHz)     & 300    & 300         & 500         \\
		Sweep time ($\mu$s)           & 20     & 20          & 20          \\
		Sweep direction           & Up     & Down        & Up          \\
		PRT ($\mu$s)                  & 30     & 43          & 61          \\ 
		Sampling frequency (MHz)  & 20     & -           & -          \\ \hline
	\end{tabular}
\end{table}

\section{Experiment Results for Real Scenario} \label{sec:RealExperi}
In this experiment, real sceario data are collected for verifying the effectiveness of the proposed method. Three 77GHz millimeter-wave radars, from \href{https://www.muniutech.cn/vehicle?category_id=9}{Muniu Technology Co., Ltd.}, are used for data collection. Among these radars, one is used as a victim radar, the other two are used as interferer radars. Experimental data of the victim radar is recorded. The device positions in the scenario are shown in Fig.~\ref{fig_8} (a). The interferer radars were set on the left and the right sides relative to the LOS of the victim radar. The distance from the victim radar to the interferer radar one and the interferer radar two were 20m and 30m, respectively. Radar configurations are shown in Table~\ref{tab:table3}. Considering the ease of implementation on actual signal processing chips, all the radars were set to have the same sweep time but different sweep bandwidth to generate LFM signals with different slopes. The victim radar was configured as the up-frequency modulation mode with the sweep bandwidth of 300MHz. Interferer radar one was configured as the down-frequency modulation mode with the sweep bandwidth of 300MHz, and interferer radar two was configured as the up-frequency modulation mode with the sweep bandwidth of 500MHz. The pulse repetition time (PRT) of the radars was set to be different to increase the probability of the mutual interference. For a single chirp, the number of sampling points is 400. A window length of 32 is used for the STFT and a step between the sliding windows is set to 4. A signal in each window is applied to 128-point FFT to obtain the TF spectrogram. For a chirp sequence, a total of 128 chirps are included as a coherent processing unit for a RD response.

\subsection{Stationary Target Experiment} \label{sec:RealExperi_cornerRef}
A corner reflector was placed at 20m in front of the victim radar to simulate a typical strong target. The time domain signal and the TF spectrogram of the received signal are shown in Fig.~\ref{fig_9}. It can be seen that two forms of interference related to the interferer radars are observed in the time domain as shown in Fig.~\ref{fig_9} (a). The interference with short duration is introduced by the interferer radar one, and the one with long duration is introduced by the interferer radar two. Fig.~\ref{fig_9} (b) shows the TF features of the received signal that includes LFM-like interference and the single frequency-like target echo.
\begin{figure}[ht]
	\centering   %  居中
	% 插入图片
\includegraphics[width=2.8in]{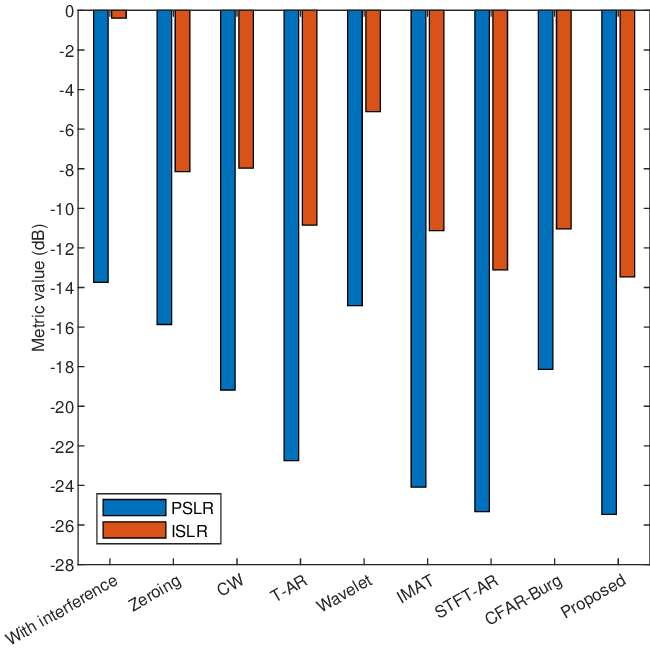}
	% 图片具体信息
	
	%\includegraphics[width=3.5in]{fig3}
	\caption{The PSLR and The ISLR of the recovered signal for the corner reflector experiment.}
	
	\label{fig_11}
\end{figure}
\begin{figure}[!h]
	\centering   %  居中
	% 插入图片
	% subfig插入子图
	\subfloat[]{\includegraphics[width=1.65in]{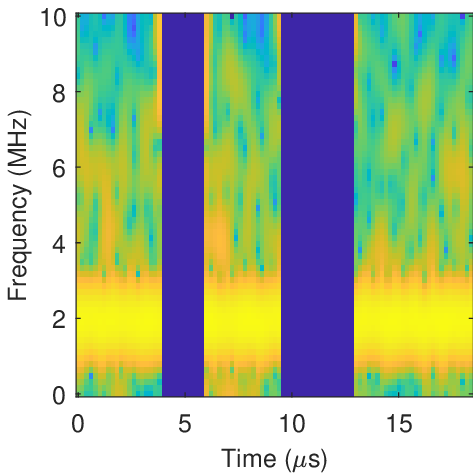}}\hspace{.1in}
	\subfloat[]{\includegraphics[width=1.65in]{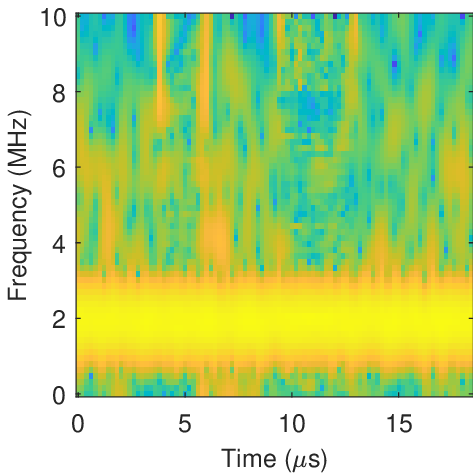}}\\
	\vspace{-0.1in}
	\subfloat[]{\includegraphics[width=1.65in]{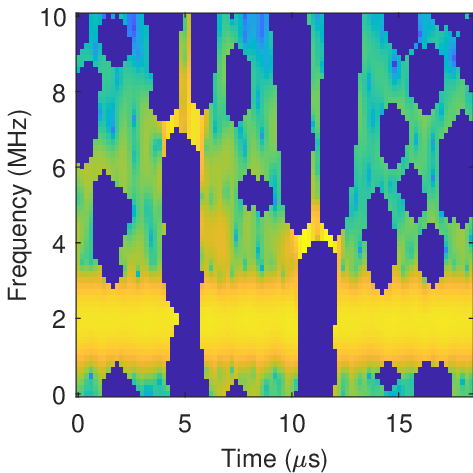}}\hspace{.1in}
	\subfloat[]{\includegraphics[width=1.65in]{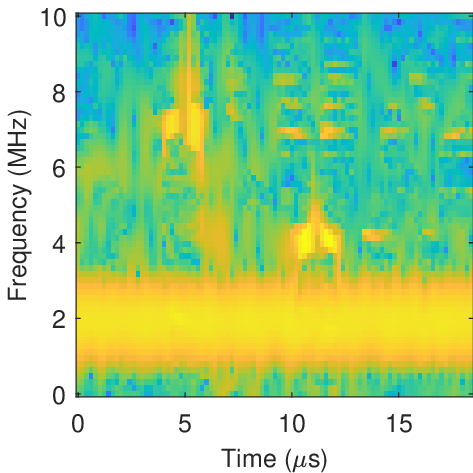}}\\
	\vspace{-0.1in}
	\subfloat[]{\includegraphics[width=1.65in]{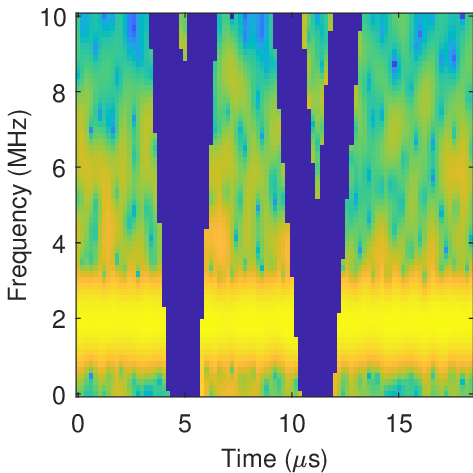}}\hspace{.1in}
	\subfloat[]{\includegraphics[width=1.65in]{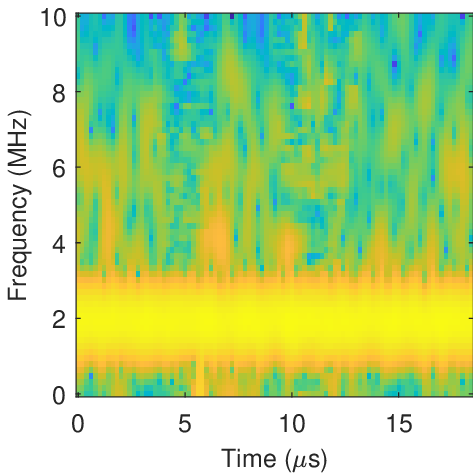}}
	% 图片具体信息
	
	%\includegraphics[width=3.5in]{fig3}
	\caption{Interference detection and mitigation performance in the TF domain for the corner reflector experiment. (a) Interference removal area for the STFT-AR method, (b) Interference mitigation by the STFT-AR method, (c) Interference detection by CFAR, (d) Interference mitigation by the CFAR-Burg method, (e) Interference detection by the Hough transform, (f) Interference mitigation by the proposed method.}
	\label{fig_12}
\end{figure}

The range profile after interference mitigation for all the tested methods are shown in Fig.~\ref{fig_10}. Overall, the TF domain methods achieve better interference mitigation performance in experiment data than that of the time domain methods. The TF domain methods result in a lower noise floor level about -30dB near the target, where the time domain methods have the noise floor level about -25dB.

Fig.~\ref{fig_11} shows the PSLR and the ISLR of the corner reflector in range profile, the quantitative results can be seen that the TF domain methods are superior to the time domain methods in both the PSLR and the ISLR, except the PSLR of the CFAR-Burg method. Unlike the PSLR, which reflects the side lobe level at a certain point, the ISLR reflects the average value of the side lobe level within a certain range, so it is more accurate for evaluating the interference suppression performance. Therefore, even though the PSLR of the CFAR-Burg method is higher than some time domain methods, it still achieved better performance in overall. For the TF domain methods, the proposed method achieves the best performance on the PSLR and the ISLR.

The effects before and after interference localization and mitigation for the three TF domain methods are shown in Fig.~\ref{fig_12}. For the STFT-AR method, the interference location in this experiment is manually marked since no method for interference detection is given in the original literature \cite{16}, so it achieves a better interference mitigation effect. However, the performance of the STFT-AR method will be lower than the results in this paper in practice, since the interference detection are not as accurate as manual marking.

For the CFAR-Burg method, two factors affect the interference location detecting as shown in Fig.~\ref{fig_12} (c). One factor is that two adjacent interference signals in the same frequency slice will raise detection thresholds for each other during the CFAR detecting, causing interference missed detection at certain frequencies. The other factor is false alarms caused by the low SNR, which lead to the loss of useful information in the TF domain. Compared with the STFT-AR and the CFAR-Burg methods, the proposed method detects the interference location more accurately in the measured data due to the utilization of the structural information in the TF domain, therefore the best interference mitigation effect is achieved.
\begin{figure*}[!t]
	\centering   %  居中
	\subfloat[]{\includegraphics[width=1.75in]{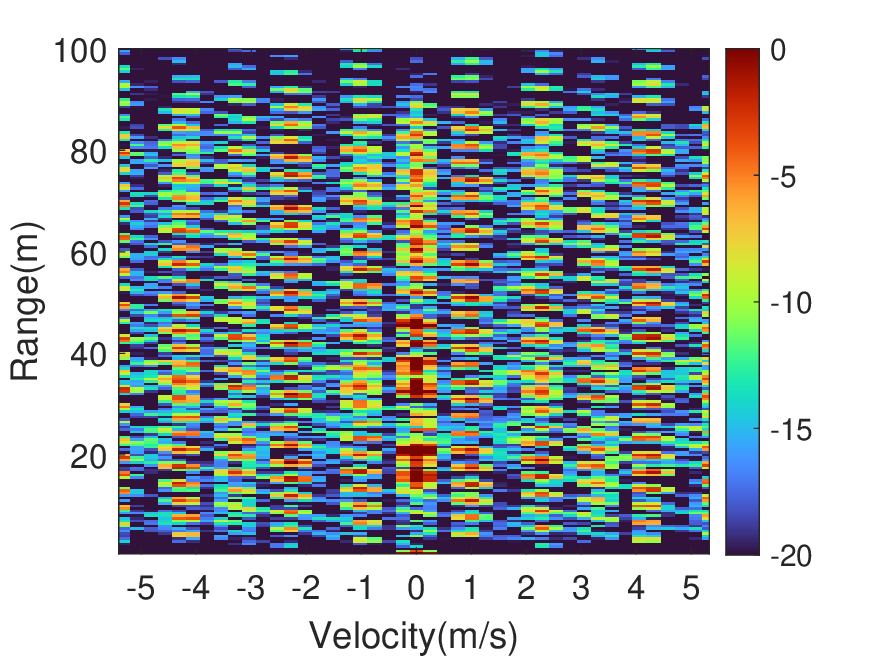}}
	\subfloat[]{\includegraphics[width=1.75in]{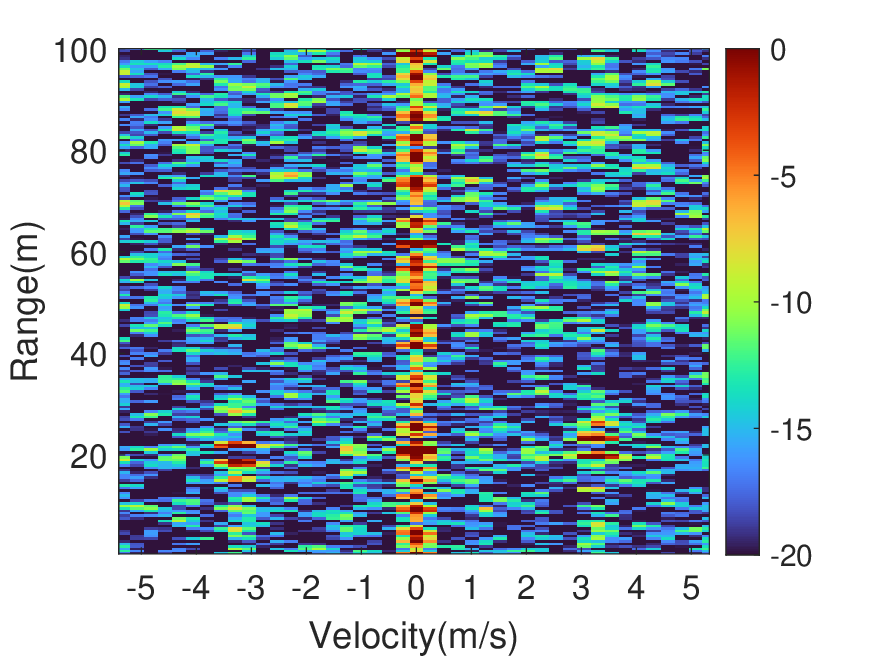}}
	\subfloat[]{\includegraphics[width=1.75in]{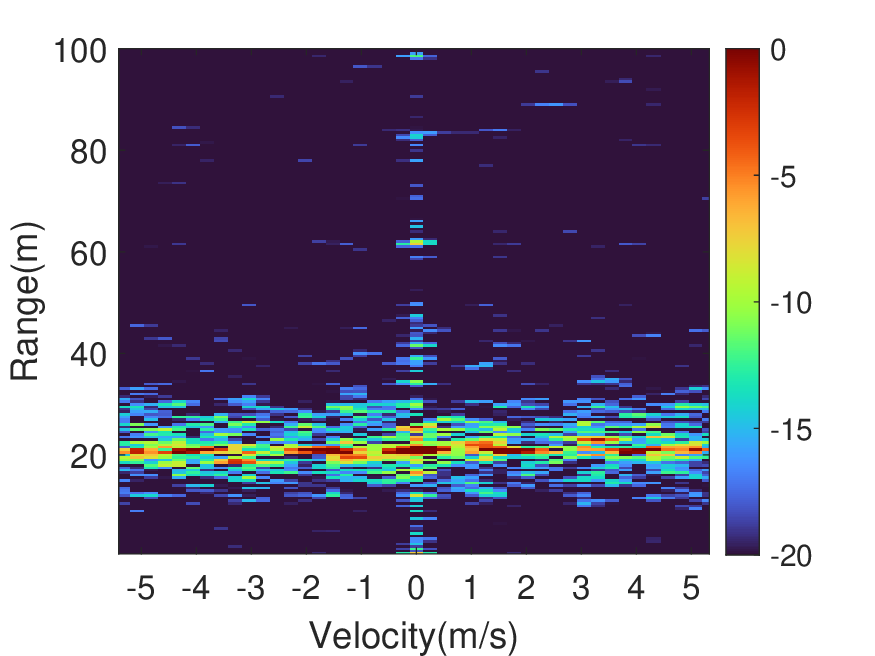}}
	\subfloat[]{\includegraphics[width=1.75in]{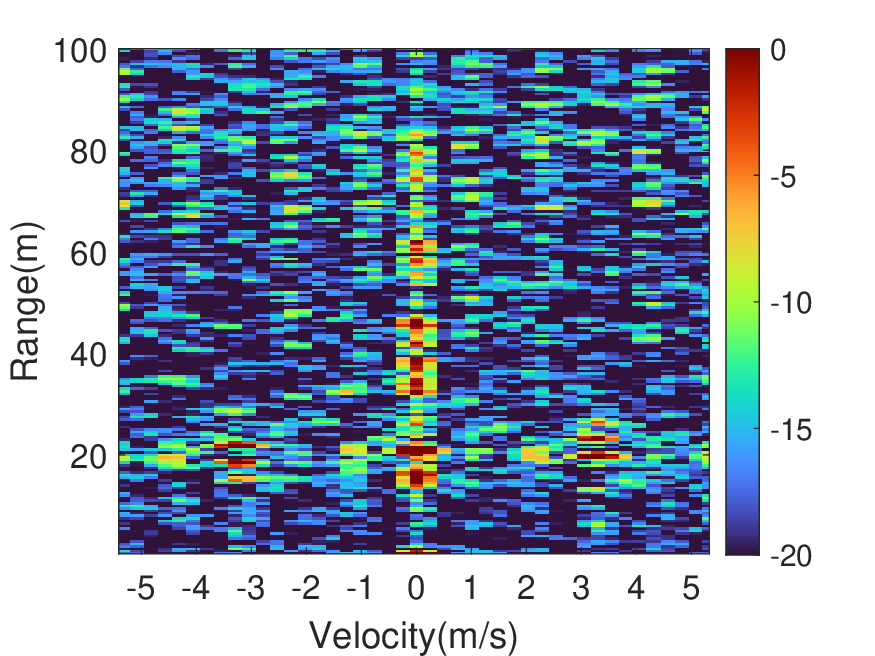}}\\
	\vspace{-0.1in}
	\subfloat[]{\includegraphics[width=1.75in]{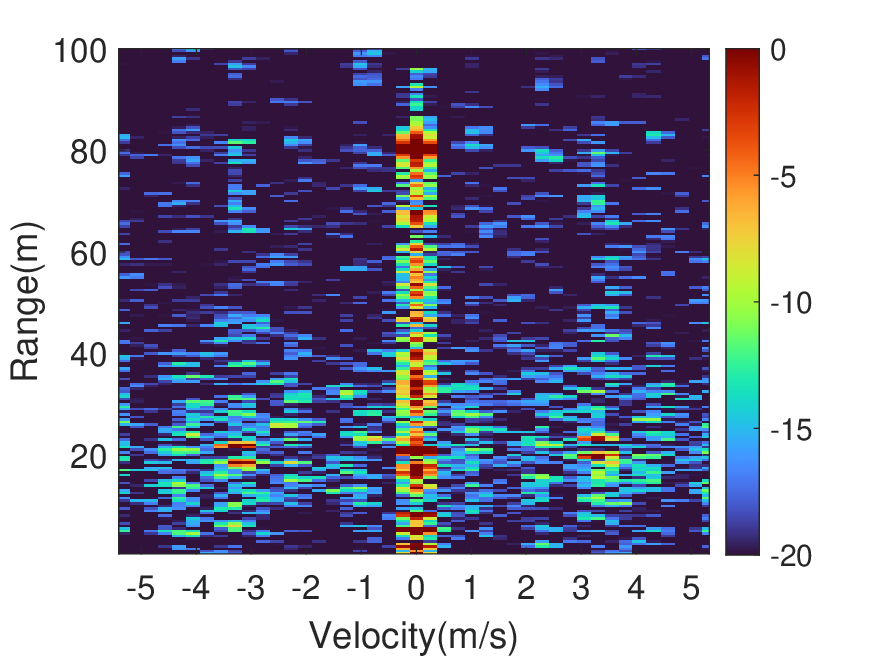}}
	\subfloat[]{\includegraphics[width=1.75in]{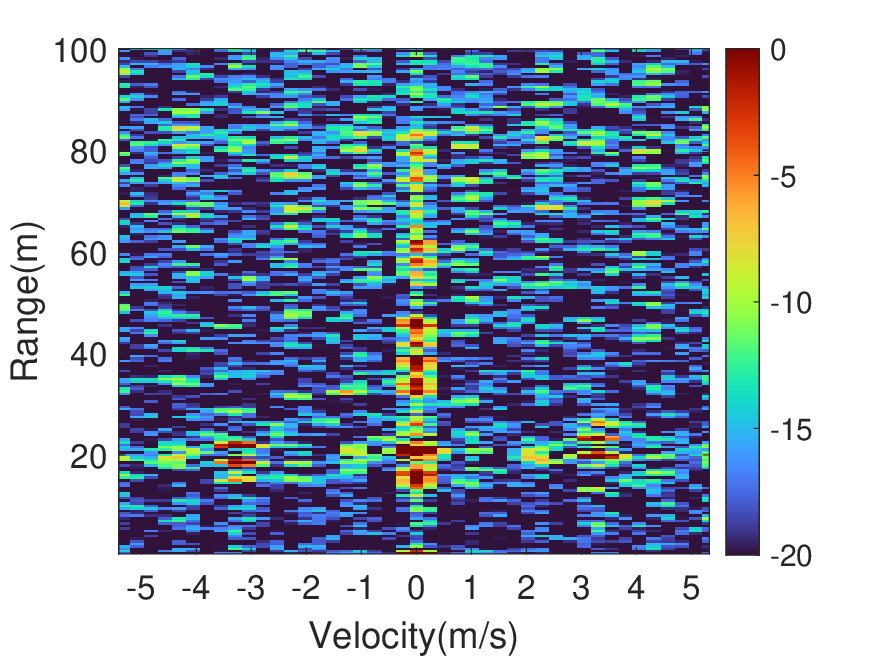}}
	\subfloat[]{\includegraphics[width=1.75in]{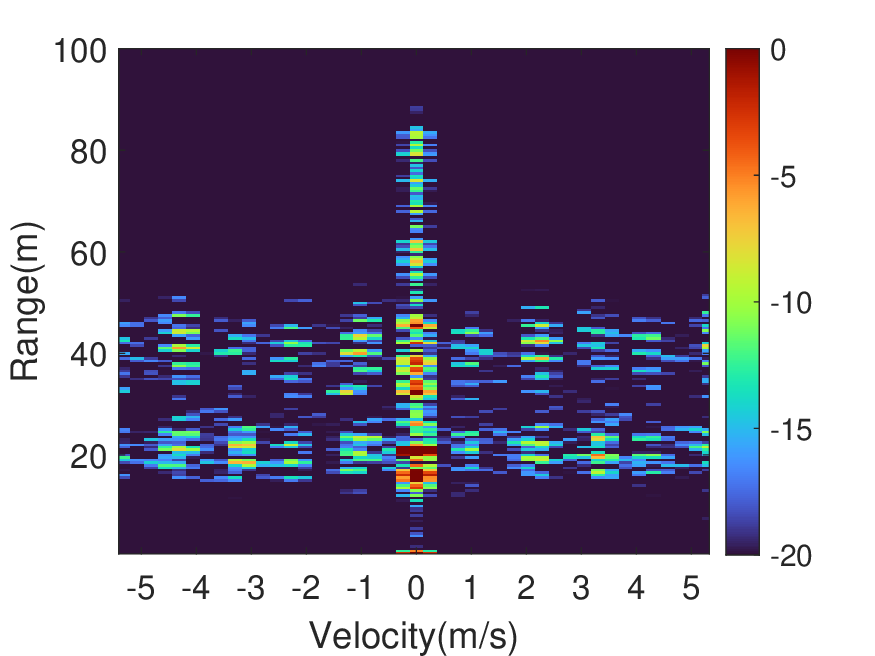}}
	\subfloat[]{\includegraphics[width=1.75in]{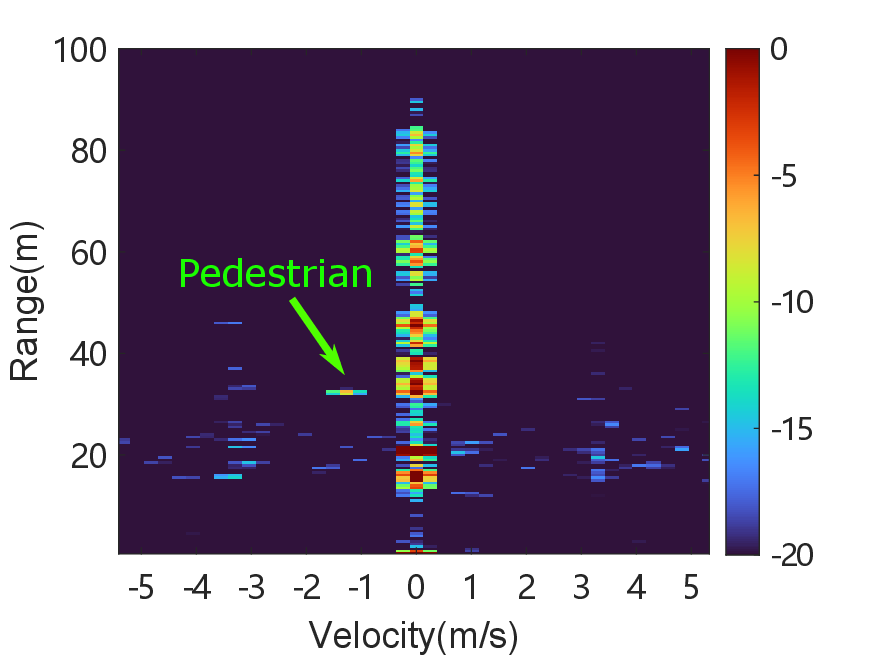}}
	\caption{Zoomed in RD responses before and after interference mitigation for the pedestrian experiment by the tested methods. (a) RD response with interference, (b) Zeroing, (c) CW, (d) T-AR, (e) Wavelet, (f) IMAT, (g) CFAR-Burg, (h) The proposed method.}
	\label{fig_realRd}
\end{figure*}

\subsection{Pedestrian Experiment} \label{sec:RealExperi_Pedestrian}
A pedestrian walking back and forth at a range of 30m to 40m from the victim radar is used to evaluate the performance of the tested methods in the interference scenario as shown in Fig.~\ref{fig_8} (b). The RD responses are obtained for a coherent processing unit, i.e., the 128 chirps, with 400 time sampling points per chirp, by performing the range FFT and the Doppler FFT with Hamming window \cite{oppenheim2001discrete}. The processing flow of the RD responses is similar to the simulation implemented in Section~\ref{sec:SimExperi_MvTar}. For the pedestrian, the RD responses obtained by the tested methods are shown in Fig.~\ref{fig_realRd}.

In the pedestrian experiment, the presence of a strong target such as the corner reflector makes the difference in power between the target echo and the interference no longer significant, which increases the difficulty of interference detection and localization. In this case, the interference mitigation by the time domain methods, despite the improvement, is not sufficient to achieve the required SNR for pedestrian detecting. Therefore, the time domain methods can not provide an effective measurement of the pedestrian’s range and velocity in the RD response as shown in Fig.~\ref{fig_realRd} (b) to Fig.~\ref{fig_realRd} (f). For the CFAR-Burg method, the pedestrian just barely appeared in the RD response after interference mitigation due to the existence of false alarms and interference missed detection problems as shown in Fig.~\ref{fig_realRd} (g). The interference missed detection is a failure point of the CA-CFAR approach in interference-dense scenarios. When the locations of multiple interference are close to each other in time as shown in Fig.~\ref{fig_12} (c), it causes the CFAR detector to overestimate noise levels, which raises the detection threshold and lead to the missed detection. The same phenomenon of the interference missed detection can be seen in the original literature of the CFAR-Burg method [34]. For the proposed method, the interference can be better mitigated in the pedestrian experiment, resulting in the correct detection of the pedestrian’s range and velocity as shown in Fig.~\ref{fig_realRd} (h).
\begin{table}[ht]
	\caption{Aogorithmic Runtime for the Tested Methods \label{tab:Runtime}}
	\centering
	\begin{tabular*}{\hsize}{@{\extracolsep{\fill}}lcr}
		\hline
		Method name                        & Method class  & Runtime ($\text{ms}$)       \\ \hline
		Zeroing \cite{nozawa2017anti}                           & Time domain   & 4.8    \\
		CW \cite{nozawa2017anti}                                & Time domain   & 4.8    \\
		T-AR \cite{15}                               & Time domain   & 14.4          \\
		IMAT \cite{bechter2017automotive}                               & Time domain   & 5.2       \\
		Wavelet \cite{13}                            & Time domain   & 8.1            \\
		STFT-AR \cite{16}                         & TF domain     & 112.4                    \\
		CFAR-Burg \cite{14}                         & TF domain     & 490.8         \\
		\textbf{Proposed} & \textbf{TF domain}     & \textbf{560.2} \\ \hline
	\end{tabular*}
\end{table}

\subsection{Algorithmic Runtime Analysis} \label{sec:RealExperi_Runtime}
The runtime results of the tested methods are evaluated by using the data from the corner reflector experiment in Section~\ref{sec:RealExperi_cornerRef}. For a chirp signal with interference, the interference mitigation process from the tested methods is applied and the corresponding runtime is recorded. The MATLAB version used in the experiment is R2021a and the computer configurations are AMD Ryzen 7 5800H CPU and 16GB DDR4 3200MHz RAM. The runtime results of the tested methods are shown in Table~\ref{tab:Runtime}.

Overall, the TF domain methods have longer runtime than that of the time domain methods because they expand an one-dimension time signal into a two-dimension TF spectrogram, and implement interference mitigation processes in the TF domain. For the STFT-AR method, since there is no interference detection and localization step, its runtime is mainly consumed in the STFT and the signal reconstruction process. For the CFAR-Burg and the proposed methods, due to the presence of the interference detection and localization steps, their runtime have a large increase compared with the STFT-AR method, which indicates the interference detection and localization in the TF domain is the most time-consuming parts of the TF methods. In addition, the Hough transform used in the proposed method detects lines by a search process in a two-dimensional parameter space and therefore has the longest algorithm running time. However, since the search grids of the Hough parameter space are independent of each other, parallel processing can be considered for reducing the runtime in practice.

\section{Conclusions}  \label{sec:Conclu}
In this paper, the mutual interference of automotive radars in the TF domain is analyzed. Based on the linear characteristic of the interference in the TF domain, a power-weighted Hough transform interference detection approach is proposed, and then the AR model based predicting is used for interference mitigation. Compared with the existing interference mitigation methods implemented in the time domain, the proposed method has the ability to locate the interference more accurately in the TF domain, and retains more useful signals in the interference mitigation process. Compared with the STFT-AR and the CFAR-Burg methods implemented in the TF domain, the proposed method accumulates interference power based on structural information for improving detection and location performance. As a result, the target echo can be recovered more accurately and robustly under low SNR conditions.

% \newpage

\bibliographystyle{IEEEtran}
\bibliography{InterferenceMitigationRef}

\begin{IEEEbiography}[{\includegraphics[width=1in,height=1.25in,clip,keepaspectratio]{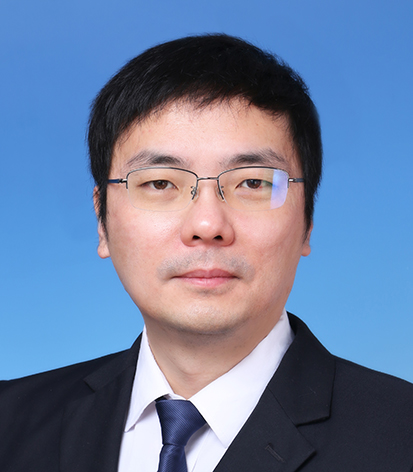}}]{Yanbing Li} (M'22) received the M.S. and Ph.D. degrees in signal and information processing	from Xidian University, Xi'an, China, in 2009 and 2013, respectively. 
	
	He is now an associate professor with the School of Electronic and Information Engineering, Beijing Jiaotong University, Beijing, China. His research interests include radar system design, radar signal processing, radar target recognition, and the applications of radar sensing techniques in autonomous driving, intelligent transportation and internet of things.
\end{IEEEbiography}

\begin{IEEEbiography}[{\includegraphics[width=1in,height=1.25in,clip,keepaspectratio]{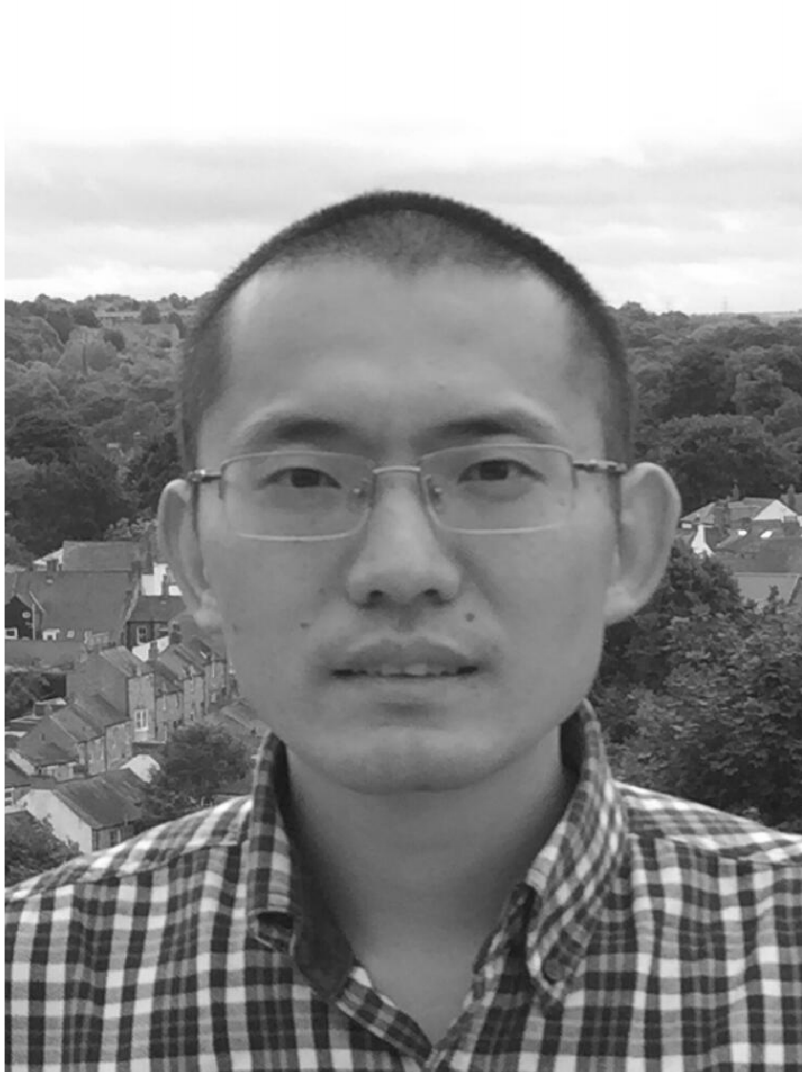}}]{Weichuan Zhang} received the M.S. degree in
	signal and information processing from the
	Southwest Jiaotong University in China and the
	Ph.D. degree in signal and information processing
	in National Lab of Radar Signal Processing,
	Xidian University, China. He is a research
	fellow at Griffith University, QLD, Australia.
	His research interests include computer vision,
	image analysis, and pattern recognition. He is a
	member of the IEEE.
\end{IEEEbiography}

\begin{IEEEbiography}[{\includegraphics[width=1in,height=1.25in,clip,keepaspectratio]{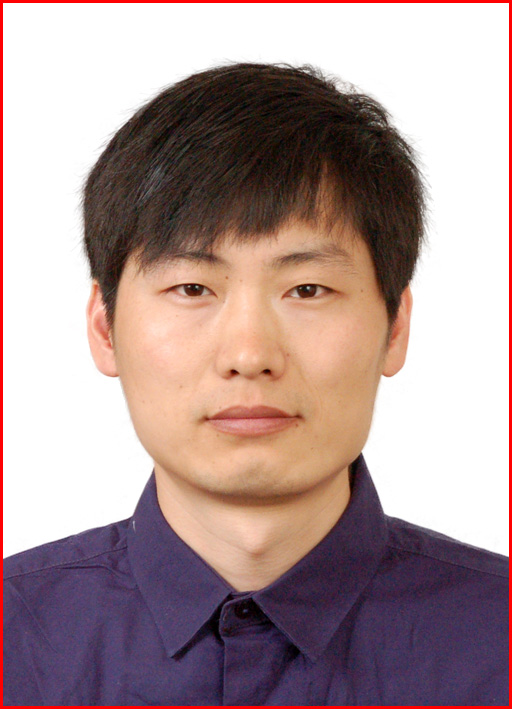}}]{Lianying Ji} received the B.S. degree from the Dalian Maritime University, in 2004, and the Ph.D. degree from the Beijing Institute of Technology in 2009.
	
	He has been with the School of Electronic, Electrical and Communication Engineering, University of Chinese Academy of Sciences, since 2009. During 2010, He was visitor researcher in China-Singapore Institute of Digital Media, Singapore. He is now the CTO of Beijing Muniu Linghang Technology Company. His technical contributions have been in the area of both biomedical information processing and  mmWave radar signal processing.
\end{IEEEbiography}

\end{document}